\documentclass[fleqn,usenatbib]{mnras}

\usepackage{newtxtext,newtxmath}

\usepackage[T1]{fontenc}

\DeclareRobustCommand{\VAN}[3]{#2}
\let\VANthebibliography\thebibliography
\def\thebibliography{\DeclareRobustCommand{\VAN}[3]{##3}\VANthebibliography}

\usepackage{graphicx}	
\usepackage{amsmath}	
\usepackage{amssymb}	

\title[Braking indices of radio pulsars]{Braking indices of young radio pulsars: theoretical perspective}

\author[A.P. Igoshev \& S.B. Popov]{
Andrei P. Igoshev,$^{1}$\thanks{E-mail: ignotur@gmail.com}
Sergei B. Popov,$^{2,3}$
\\
$^{1}$Department of Applied Mathematics, University of Leeds, Leeds LS2 9JT , UK\\
$^{2}$Sternberg Astronomical Institute, Lomonosov Moscow State University, Universitetsky prospekt 13, 119234, Moscow, Russia\\
$^3$ Higher School of Economics, Department of Physics,
 Moscow, 101000, Russia
}

\date{Accepted XX. Received YYY; in original form ZZZ}

\pubyear{2015}

\begin{document}
\label{firstpage}
\pagerange{\pageref{firstpage}--\pageref{lastpage}}
\maketitle

\begin{abstract}
Recently, Parthsarathy et al.  analysed long-term timing observations of 85 young radio pulsars. They found that 11
objects have braking indices ranging $\sim 10-100$, far from the classical value $n=3$. They also noted a mild correlation between measured value of $n$ and characteristic age of a radio pulsar. In this article we systematically analyse possible physical origin of large braking indices. We find that a small fraction of these measurements could be caused by gravitational acceleration from an unseen ultra-wide companion of a pulsar or by precession. Remaining braking indices cannot be explained neither by pulsar obliquity angle evolution, nor by complex high-order multipole structure of the poloidal magnetic field. The most plausible explanation is a decay of the poloidal dipole magnetic field which operates on a time scale $\sim 10^4-10^5$~years in some young objects, but has significantly longer time scale in other radio pulsars. This decay can explain both amplitude of measured $n$ and some correlation between $n$ and characteristic age.
The decay can be caused by either enhanced 
 crystal impurities in the crust of some isolated radio pulsars, or more likely, by enhanced 
resistivity related to electron scattering off phonons  due to slow cooling of low-mass neutron stars. 
If this effect is indeed the main cause of the rapid magnetic field decay manifesting as large braking indices, we predict that pulsars with large braking indices are hotter in comparison to those with $n\approx 3$.
\end{abstract}




\section{Introduction}

Rotational evolution of young radio pulsars is usually described in a simplistic way using so-called magneto-dipole formula \citep{magnetodipole}. This description is a zeroth-order approximation at best because it lacks multiple physical effects: (1) a radio pulsar is not a point source (see \citealt{petri_multipoles} for details), (2) angle between the magnetic dipole  and the rotational axes (so-called obliquity angle $\alpha$) seems to evolve with time \citep{tauris1998}, (3) the pulsar magnetosphere is not in vacuum state  and contains charges, so the slowing down proceeds mainly due to emission of pulsar wind \citep{philippov2014}. Moreover, the poloidal magnetic field, which causes this slowing down cannot be constant due to finite crust conductivity of the neutron star (NS) and should evolve over time.

In response to challenges based on observational data, many models with evolving magnetic field (e.g., see reviews and references in \citealt{cumming2004, 2018A&G....59e5.37G, 2019LRCA....5....3P}) and angle \citep{philippov2014, 2017MNRAS.472.2142D} were developed, as well as several modifications for the equation describing rotational energy losses were proposed (e.g., \citealt{1993ppm..book.....B, 2018PhyU...61..353B} and references therein). These scenarios predict different timing evolution for single radio pulsars. Precise measurements of spin period ($P$) and its time derivatives ($\dot P, \ddot P$)  provide one of the best tests for various theoretical scenarios.  

Recently, \cite{2019MNRAS.489.3810P,listBrakingInd} measured braking indices, $n=2-(P \ddot P)/(\dot P^2)$, for 19 young radio pulsars. 
\cite{listBrakingInd} argued that this result is robust against uncertainties in the irregular (stochastic) component of pulsars' timing noise and reflects a regular pulsar evolution at least at the timescales of tens of years.
The obtained values for eleven of them are in the range $\sim 10$~--~$100$, two are negative, two are above $100$ and only four are compatible with the classical value $n=3$ predicted by the magneto-dipole formula for constant field and obliquity angle. 
 The original sample of young radio pulsars with decade-long timing measurements used by \cite{2019MNRAS.489.3810P} included 85 radio pulsars.
For most of them only upper limits  on the braking indices (in the range $n\lesssim10$~--~$2000$) were obtained.

\cite{listBrakingInd} mentioned several  hypotheses to explain large values of the braking index.
In the present article we test these hypotheses and aim at identifying the physical cause for these measurements.
First, we examine what fraction of the sample presented by \cite{2019MNRAS.489.3810P} could be in fact wide binaries with orbital periods of tens or hundreds years. Second, we systematically analyse physical phenomena occurring in truly isolated radio pulsars which affect the process of slowing down and increase the value of braking index. Such phenomena are the magnetic field decay analysed in Section~\ref{s:mfd}, obliquity angle evolution considered in Section~\ref{s:obliquity}, rigid body precession calculated in Section~\ref{s:precession} or complicated multipolar configuration of the poloidal magnetic field discussed in Section~\ref{s:nondipole}.  We conclude that only magnetic field decay provides plausible explanation for the majority of large braking indices.


\section{Wide binaries}
\label{s:wide_binaries}
A large fraction 
of NS progenitors (massive OB stars) are expected to be born 
in binaries \citep{moe2017}. Many of these binaries are disrupted at the moment of supernova explosion due to sudden mass loss and pulsar natal kick \citep{Hills1983, natalkickI, natalkickII}.

\cite{2019MNRAS.485.5394K} analysed population of compact objects in supernova remnants to probe the fraction of binaries among them. These authors provide the following estimates: $<10$~per~cent of non-interacting binaries with NSs survive after core-collapse SNae, and so do $\lesssim 10$~per~cent of interacting ones. In the case of objects under study, we are definitely not dealing with very close binaries, as otherwise the orbital motion could be directly recognized via pulsar timing \citep{2019MNRAS.489.3810P}. Thus, we are possibly left with wide pairs having orbital periods $\gtrsim$~tens of years. 

NSs born with weak natal kicks could stay bound even in wide binaries with relatively massive companions. 
If an orbital period exceeds 50-100 years, periodic fluctuations are not seen in the timing measurements of radio pulsar because undisturbed timing series continues for a few decades at best for majority of radio pulsars. Also, many systems might be highly eccentric after an SN explosion, thus a pulsar might spend a long time in apastron, where orbital velocity is very low.

Unseen as periodic modulation in time series, the companion gravity still accelerates the pulsar contributing to measured $\ddot P$ and increasing or decreasing the braking index.
According to \cite{2019MNRAS.485.5394K} we are left with small (but non-negligible) fraction of such systems, few per cent at most.  
\cite{2019MNRAS.489.3810P} discussed the possibility that PSR J0857–4424 is a member of a wide binary system with a NS or a WD as a companion. 
For two pulsars in the sample (PSR  J1637–4642, PSR J1830–105) presence of planetary companions have been discussed \citep{listBrakingInd}, but detailed analysis did not confirm this hypothesis. 

\begin{figure}
    \centering
    \includegraphics[width=1.0\linewidth]{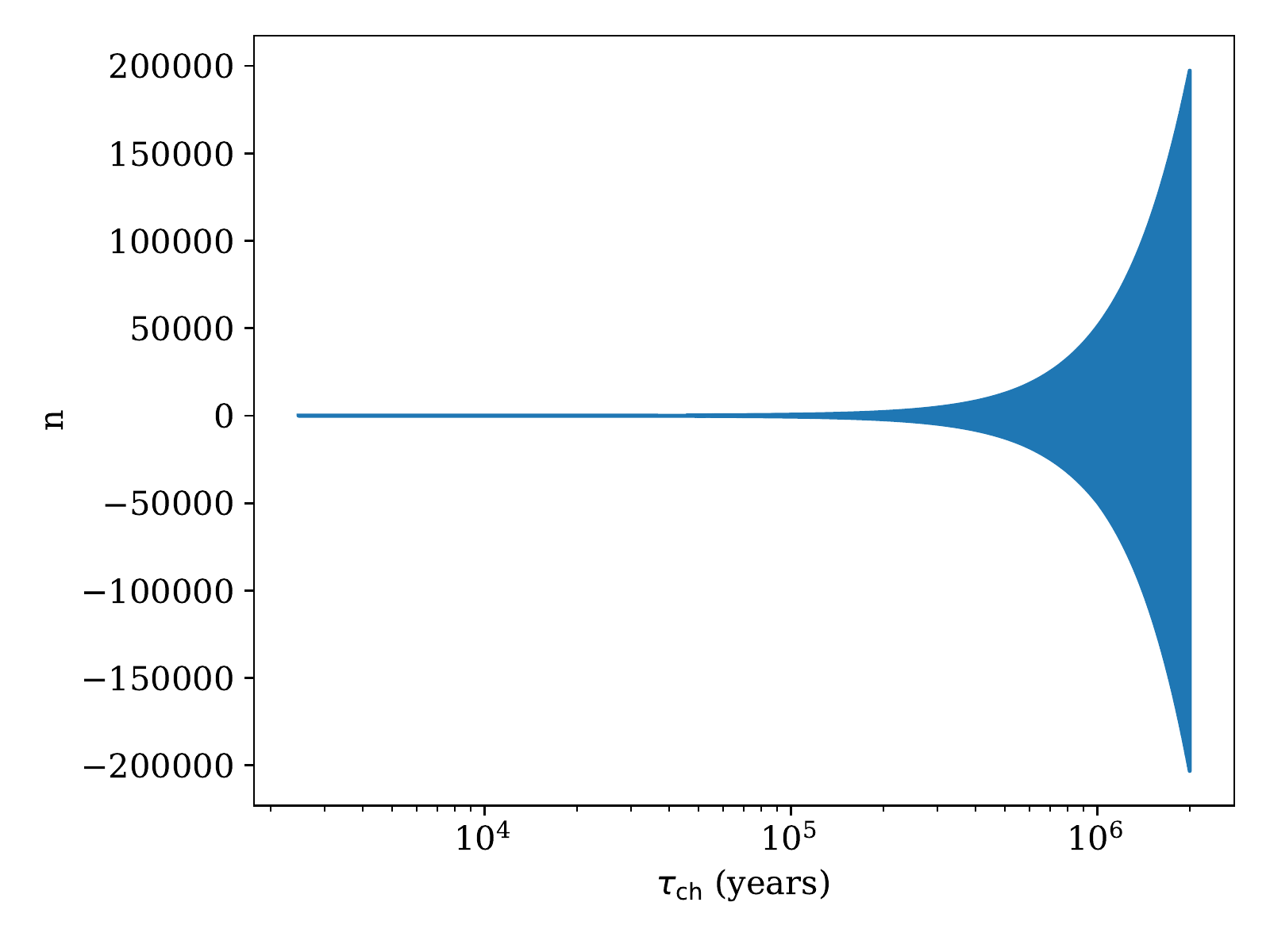}
    \caption{The expected oscillations of braking index $n$ for mass of possible unseen companion $M_c = 1$~M$_\odot$ to PSR J0857-4424 as a function of spin-down age $\tau_\mathrm{ch}$. Blue region is filled with multiple oscillations braking index caused by motion of unseen companion. Braking index oscillates with assumed orbital period of $130$~years. 
    }
    \label{f:mixed}
\end{figure}

Another way of looking at the \cite{2019MNRAS.489.3810P} data is to count number of objects with positive and negative braking indices. If all large braking indices are caused by presence of a companion on a wide orbit, we would expect  random orientation of this companion's orbit in respect to the line of sight and random orbital phase, thus the gravitational acceleration (and braking index) should be positive in approximately half of the cases ($p=0.5$) and negative in the other half (see eq. 19 in \citealt{listBrakingInd} which depends on the orbital phase $\phi$ as $\cos\phi$). We plot oscillations of braking index for parameters of PSR J0857-4424 estimated by \cite{listBrakingInd} (mass of unseen companion $M_c = 1$~M$_\odot$, orbital period $130$~years, inclination $i=30^\circ$). As clearly seen from the figure the braking index is indeed negative half of the time.


We can quantify a probability of finding $k=2$ negative braking indices in $n=19$ measurements. This is done using the binomial distribution. 
The resulting probability is:
\begin{equation}
f(k,n,p) = \frac{n!}{k! (n-k)!} p^k (1-p)^{n-k}\approx 3\times 10^{-4},    
\end{equation}
which is very low. Therefore, it is highly unlikely that all non-standard braking indices could by explained by presence of massive unseen companion on a wide orbit.

Additionally, to even better constrain possible fraction of wide companions, we search for optical counterparts to 19 pulsars with measured braking indices using the second Gaia data release \citep{gaiaI, gaiaII}. We search for all stars located at an angular separation up to $7\arcsec$ from the catalogue position of the radio pulsar. We find that 12 out of 19 pulsars with measured  braking indices have an optical counterpart in the Gaia catalogue. Most of these stars have G band magnitude $\sim 18$--20 and  angular separation  $\approx 2\arcsec$. Neither  parallax measurements for such dim stars are precise nor many pulsars have  parallax measurement, so further work is required to check if any of these optical counterparts are in fact associated with any pulsar. 

With the modern data it is possible to statistically check if a part or all of these counterparts can be a simple projection  of otherwise physically unrelated sources. To do so we add half a degree to the right ascensions of all real pulsars and search for counterparts to these ``fake'' sources. We find 10 counterparts. If we subtract half a degree from the right ascension of the real pulsars and repeat our search for counterparts we find 6 counterparts. It means that a few counterparts to real pulsars could be the pulsars themselves or their distant companions visible in optical band. Of course the Gaia could still miss multiple sources fainter than $20$th magnitude.

It is important to discuss the case of PSR J0857-4424 as it has the largest measured braking index of $n=2890\pm 30$. In our search for optical counterparts we find a star Gaia DR2 5331775184393659264 located at angular separation of 0\farcs44 from the pulsar sky location. The parallax of this star is $3.28\pm 0.40$~mas, which results in distance estimate of $\approx 300$~pc while the pulsar J0857-4424 seems to be located at distance $d_\mathrm{DM}=2.83$~kpc or $d_\mathrm{DM}=1.94$~kpc using the electron density model  YMW2017 \citep{ymw2017} and NE2001 \citep{ne2001} respectively. According to \cite{ymw2017} the pulsar distance should be in range of $0.1-1.9 d_\mathrm{DM}$ in $95$~per~cent of cases, so $0.1 \times 2.83 \approx 300$~pc is compatible with the Gaia distance. In reality, the distance estimated by the dispersion measure falls more frequently outside of this wide interval \citep{deller2019}.
Unfortunately, it is impossible to conclude without further observations if the star seen in the Gaia is a pulsar itself, its companion or a completely unrelated object. A wide orbit companion to PSR J0857-4424 was discussed in \cite{listBrakingInd} with no solid conclusion. 

Overall, following these multiple lines of reasoning described above we derive that a few objects (our guess is from two to six objects) in the sample by \cite{listBrakingInd} of pulsars with large braking index could be objects with a distant companion. This fraction of $4/19\approx 21$~per~cent to some extent exceeds the stellar statistics expectations \citep{2019MNRAS.485.5394K}. Four pulsars with braking indices caused by gravitational acceleration nicely agrees with two negative braking indices and with an excess of optical counterparts seen in the Gaia data. We must conclude that remaining $\approx 9-13$ radio pulsars have large braking indices which cannot be explained by their gravitational acceleration.

\section{Magnetic field decay}
\label{s:mfd}
The process of slowing-down of NSs rotation is typically controlled by large scale poloidal magnetic field with an exception of magnetars in an active state, where the crustal motions produce magnetospheric twist; i.e. external field is not purely poloidal anymore, and contains toroidal component. The twist leads to enhanced spin-down of the NS \citep{beloborodov2009}. Such a twist is unstable and the magnetosphere tends to untwist itself quasi-steadily \citep{2017ApJ...844..133C} or through a magnetic reconnection episodes \citep{2012ApJ...754L..12P}. These reconnection events in overtwisted magnetospheres produce high-energy outbursts. Before a strong flare as the twist grows, spin-down rate  increases and X-ray spectrum becomes harder. This type of behaviour was clearly observed in the case of the hyperflare of SGR 1806-20 in 2004 (see \citealt{2008A&ARv..15..225M} and references therein). 
Such activity is  not typical for normal radio pulsars. These sources might have moderate toroidal field in the crust, but this does not significantly influence their rotational evolution. Due to these factors we consider from here on only poloidal magnetic field. 

Magnetic field decay in isolated radio pulsars can be described as an action of three distinct  processes (e.g., \citealt{IgoshevPopov2015} and references therein) driven by: (1) scatter of electrons off crystal phonons, which we refer to as ``phonon resistivity'', (2) Hall evolution, and (3)  resistivity due to crystal impurities in the crust. Correspondingly, one can introduce three timescales: $\tau_\mathrm{ph}, \tau_\mathrm{Hall}, \tau_\mathrm{Q}$.

The expected timescale for the decay due to phonon resistivity is related to cooling of NS and for the  so-called minimal scenario of thermal evolution (i.e., neglecting direct URCA processes, exotic phases, additional heating due to field decay, etc., see e.g. \citealt{2004ApJS..155..623P}; we use in our estimates a particular realization presented by \citealt{Shternin2011_cooling_curves})  the mean timescale seems to be around 400~kyr  \citep{IgoshevPopov2015}. 

The Hall time scale depends on the magnetic field of a NS as  \citep{cumming2004}:
\begin{equation}
\tau_\mathrm{Hall} = \frac{4\pi n_e e L^2}{c B_p}, \end{equation}
where  $c$ is the speed of light, $e$ is the elementary charge,  $B_p$ is the poloidal dipolar magnetic field at the pole, $n_e$ is the electron number density, and $L$ is length scale related to electric currents structure in the crust. The Hall evolution is dominant in magnetars due to their high magnetic fields.
 For a normal pulsar the Hall time scale is typically about a few Myrs.
 
The timescale of decay due to crustal impurities seems to be the longest in normal pulsars. It exceeds at least 8~Myr and is compatible with the value even larger than 20~Myr \citep{Igoshev2019}.  Note that stable long-term magnetic field decay is not clearly visible in lifelong evolution of radio pulsars \citep{2006ApJ...643..332F},\footnote{Several authors \citep[e.g.]{2004ApJ...604..775G, 2010MNRAS.401.2675P, 2015MNRAS.454..615G} came to different conclusions, as they obtained better fits with inclusion of magnetic field decay. What we want to highlight, is that the evolution of the whole population of normal radio pulsars on time scale $\sim$ few tens of Myrs can be described without significant continuous field decay distinguishable in present day observations already on a time scale $\sim$ few $\times10^5$~yrs.} which probably agrees with weak  dissipation due to impurities, since it operates along the whole life span of an NS. Therefore,  crystal impurities in the crust cannot contribute significantly to field decay in all young pulsars. 
 
 Dissipation due to phonons and  impurities can be joined together to define the Ohmic time scale:

\begin{equation}
    \tau_\mathrm{Ohm}^{-1}=\tau_\mathrm{ph}^{-1} + \tau_\mathrm{Q}^{-1}.
    \label{e:tau_ohm}
\end{equation}
The Hall process is not a dissipative one, only changing the field structure in the crust. Thus, due to this reconfiguration which results in decrease of the length scale $L$, the rate of dissipation can be significantly enhanced, as $\tau_\mathrm{Ohm}\propto L^2$ \citep{cumming2004}.

Magnetic field decay affects the first and the second period derivatives and therefore, affects both spin-down age and braking index and makes them correlate with each other. Below, we first estimate the timescale of decay and then construct a more sophisticated model comparing it particularly to the measurement presented by \cite{listBrakingInd}. 

\subsection{Simple timescale estimate}
\label{s:simple_est}

The braking index is a well-known characteristic of radio pulsars (see, e.g., \citealt{1988MNRAS.234P..57B} and references therein). It can be defined as:
\begin{equation}
n = \frac{\Omega \ddot \Omega}{\dot \Omega^2}=2 - \frac{P \ddot P}{\dot P^2}.
\label{e:brk_ind}
\end{equation}
If for a moment we assume no evolution of the obliquity angle and no changes in inertia moment, we can absorb these values in a constant. So, the pulsar evolution is described as:
\begin{equation}
P \dot P = \frac{2}{3}\beta B_p^2 
\label{e:slowing}
\end{equation}
where $\beta$ is:
\begin{equation}
\beta = \frac{\pi^2 R_\mathrm{NS}^6}{Ic^3},    
\end{equation}
where $R_\mathrm{NS}$ is the neutron star radius and $I$ is its moment of inertia. For typical parameters of a NS --- $R_\mathrm{NS} = 10$~km, $I = 10^{45}$~g~cm$^2$ --- the value of $\beta$ is equal to $\beta=3.7\times 10^{-40}$~s~G$^{-2}$.

If we combine eqs. (\ref{e:brk_ind}) and (\ref{e:slowing}) together we can get a simple estimate for braking index when the magnetic field evolves:
\begin{equation}
n = 3 - 2 \frac{P}{\dot P} \frac{\dot B_p}{B_p}. 
\label{e:brk}
\end{equation}

Note that for increasing magnetic field the braking index can become significantly smaller than 3, and even reach large negative values. This corresponds to an upward movement in the $P$~--~$\dot P$ diagram. We do not consider such scenarios here (see \citealt{2016MNRAS.462.3689I} and \citealt{cco_3d} for a description of this possibility and references to earlier papers), as we focus on data from \cite{listBrakingInd} for pulsars with $n\sim10-100$.

If we assume that the magnetic field exponentially decays due to Ohmic losses in the crust on timescale $\tau_\mathrm{Ohm}$ we obtain:
\begin{equation}
n = 3 + 2 \frac{P}{\dot P} \frac{1}{\tau_\mathrm{Ohm}}. 
\label{e:int}
\end{equation}
It is interesting to note that as $\tau_\mathrm{ch} = P / {2\dot P}$ is the spin-down age, then eq. (\ref{e:int}) is simplified even further:
\begin{equation}
n = 3 + 4 \frac{\tau_\mathrm{ch}}{\tau_\mathrm{Ohm}}.    
\end{equation}
If we take typical values for $n$ and $\tau_\mathrm{ch}$ from \cite{listBrakingInd}, we obtain:
\begin{equation}
\tau_\mathrm{Ohm} = 10^4 - 10^5\; \mathrm{yrs}.    
\end{equation}
It is important to note that despite such rapid evolution is possible in young NSs, it can not proceed for a long time, otherwise the field dissipates too quickly and the number of observed radio pulsar might be significantly reduced.

\subsection{Approximate model}

An approximate description for field evolution with the Hall term was presented by \cite{aguilera2008}:
\begin{equation}
\frac{dB_\mathrm{p}}{dt} = - \frac{B_\mathrm{p}}{\tau_\mathrm{Ohm} (T_\mathrm{crust})} - \frac{1}{B_0} \frac{B_\mathrm{p}^2}{\tau_\mathrm{Hall_0}}.  
\label{e:decay}
\end{equation}
In this equation $\tau_\mathrm{Ohm}$ is temperature dependent and $\tau_\mathrm{Hall_0}$ corresponds to the initial value of the Hall time scale which depends on the initial magnetic field: $\tau_\mathrm{Hall_0} \propto B_\mathrm{p, 0}^{-1} $. Sometimes an analytical solution of this equation $B(t)$ is given for 
$\tau_\mathrm{Ohm}$ being constant. Such solution can also be used for varying 
$\tau_\mathrm{Ohm}$, but in this case the solution should be used iteratively on small timescales. Here, instead, we numerically integrate eq. (\ref{e:decay}). Moreover, we add the Hall attractor \citep{2014PhRvL.112q1101G}, so the Hall evolution is essentially turned off after three initial Hall timescales.

As mentioned earlier, the actual $\tau_\mathrm{Ohm}$ is a combination of the following terms: (i) phonon resistivity and (ii) resistivity due to  impurities, see eq. (\ref{e:tau_ohm}).
The phonon resistivity is sensitive to the crustal temperature, $T_\mathrm{crust}$, as $\tau_\mathrm{ph} \propto \tau_\mathrm{ph, 0} / T_\mathrm{crust}^{-2}$. 
 The $\tau_\mathrm{ph, 0}$ also depends on the depth inside the NS crust where  most of the electric current flows. 

In observations of young isolated radio pulsar usually it is only possible to probe the large scale, poloidal magnetic field of NSs. However, the total magnetic field configuration can include small scale poloidal magnetic field components and toroidal crust-confined magnetic field \citep{ferrario2015, Braithwaite2004,Bilous2019}. The magnetic field configuration is formed in convective liquid which solidifies at the proto-neutron stage. So the initial conditions (exact depth of electric current) can differ even for different neutron stars. Therefore, in this work we study: (1) whole range of possible resistivities and (2) we use effective quantities instead of exact distributions across the whole crust.  

{For example,} \cite{cumming2004} suggested the following equation to estimate  $\tau_\mathrm{ph}$:
$$
\tau_\mathrm{ph} = 2.2\, \mathrm{Myr}\, \frac{\rho^{15/6}_{14}}{T_8^2} \left(\frac{Y_\mathrm{e}}{0.05}\right)^{5/3} \left(\frac{Y_\mathrm{n}}{0.8}\right)^{10/3} \left(\frac{f}{0.5}\right)^2 \left(\frac{g_{14}}{2.45}\right)^{-2}=
$$
\begin{equation}
\hspace{5cm}=\frac{\tau_\mathrm{ph, 0}}{T_8^2}    
\label{e:timescale}
\end{equation}
where $\rho_{14}$ is the density inside the NS crust where the most electric current flows in units of $10^{14}$~g~cm$^{-3}$; $T_8$ is the deep crust temperature 
in units of $10^8$~K; $Y_e$, $Y_n$ are the number fractions of electrons and neutrons, respectively; $f$ is a factor  accounting for interactions between neutrons; finally, $g_{14}$ is the local gravity in units of $10^{14}$~cm~s$^{-2}$.  
This estimate is based on the assumption that typical height for the crust layer with current is proportional to typical pressure height. Applying eq.~(\ref{e:timescale}) for density range of $2\times 10^{13}$~g~cm$^{-3}$ to $10^{14}$~g~cm$^{-3}$ we get the estimate for typical timescale of the phonon decay  range from $4\times 10^4$~years to 2.2~Myr for $T_8 = 1$. In contrast to our previous works, here we do not turn-off the phonon resistivity when a critical temperature is reached, in agreement with results by \cite{chugunov2012}. 

\begin{figure*}
    \centering
    \begin{minipage}{0.49\linewidth}
    \includegraphics[width=1.0\linewidth]{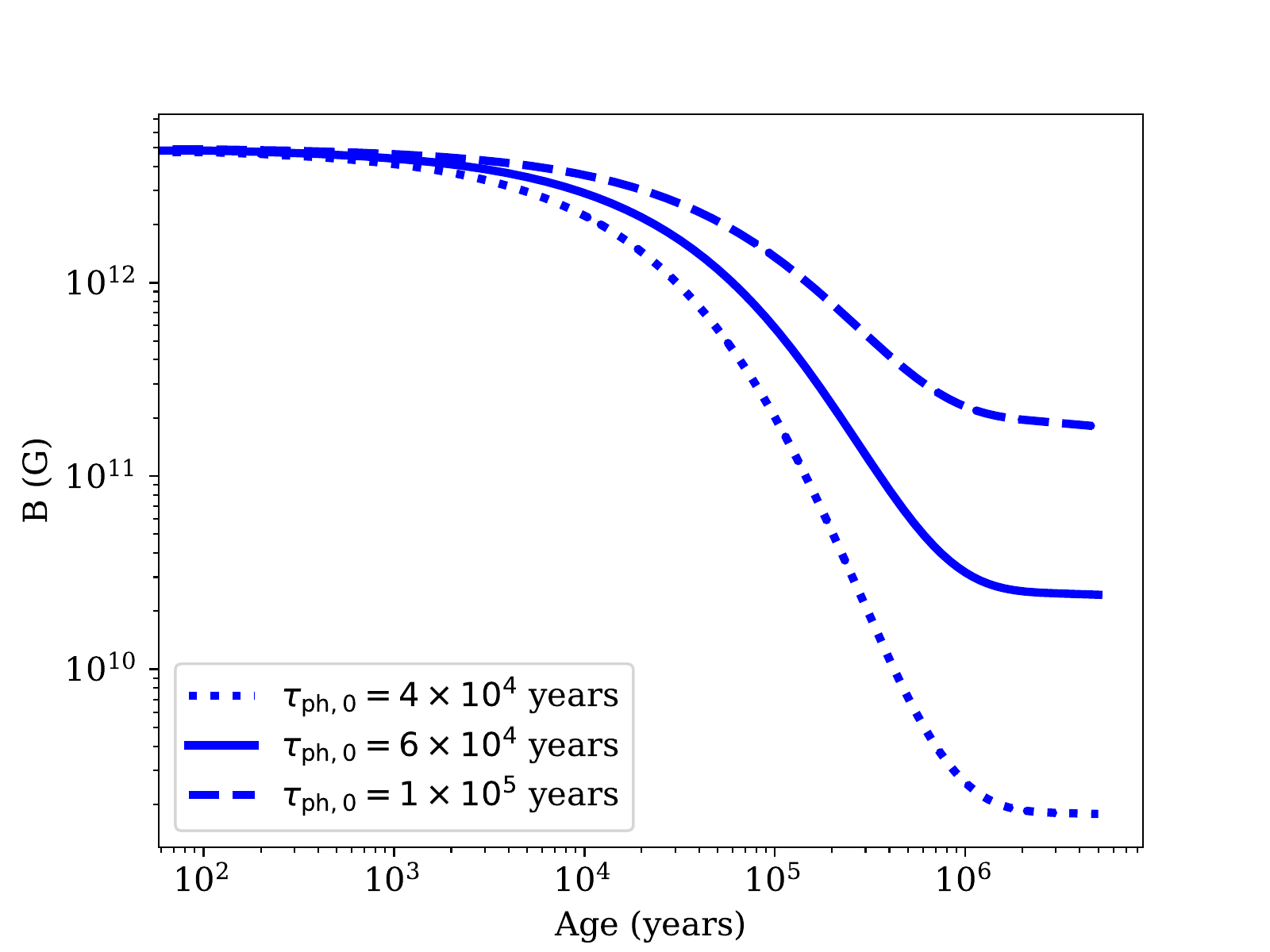}
    \end{minipage}
    \begin{minipage}{0.49\linewidth}
    \includegraphics[width=1.0\linewidth]{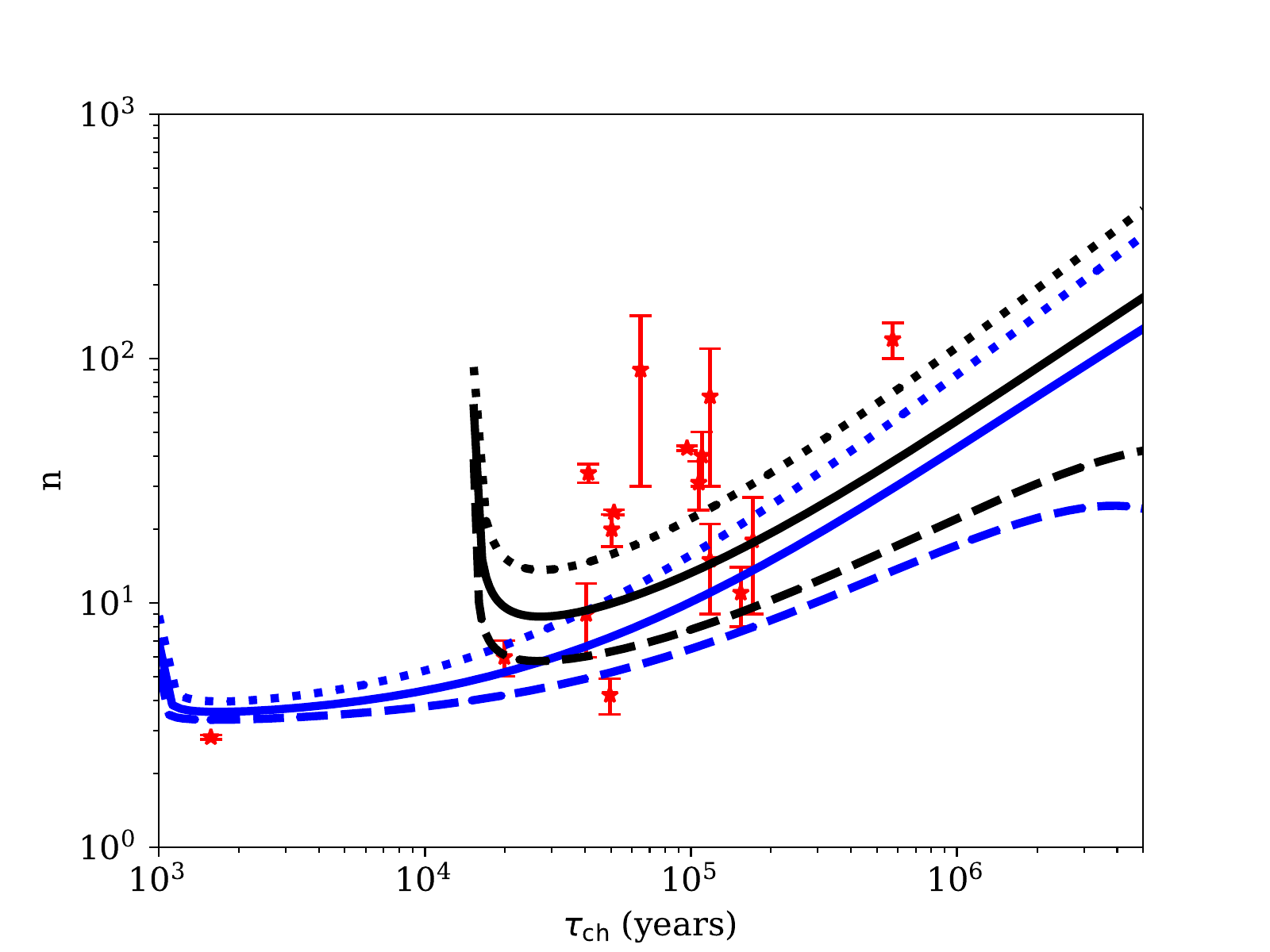}
    \end{minipage}    
    \caption{Left panel: magnetic field evolution for different value of $\tau_\mathrm{ph,0}$ for NS mass of $1.32$~M$_\odot$ and $\tau_Q = 200$~Myr. Right panel: dependence of braking index on spin-down age for different $\tau_\mathrm{ph,0}$ and initial periods. Blue lines are for $P_0 = 0.04$~s and black lines are for $P_0 = 0.16$~s. Dotted lines correspond $\tau_\mathrm{ph,0} = 4\times 10^4$~yrs, solid lines $\tau_\mathrm{ph,0} = 6\times 10^4$~yrs and dashed lines are for $\tau_\mathrm{ph,0} = 10^5$~yrs. Red dots show measured values for 16 isolated radio pulsars from \protect\cite{listBrakingInd}. }
    \label{f:mf}
\end{figure*}

The value of $\tau_\mathrm{Q}$ also depends on the crust density. Recent analysis by \cite{Igoshev2019} showed that this value is above 
8 Myr. So it cannot affect our model on timescales of $\approx 1$~Myr, unless some NSs have significantly different value of impurity, Q. At  densities around $10^{14}$~g~cm$^{-3}$ we 
can deal with the highly-resistive pasta layer \citep{Pons2013}. This layer can have impurity parameter of $Q\approx 100$ which allows the magnetic field to decay significantly by the NS age of $\approx 1$~Myr. This layer is important in explaining the absence of long-period magnetars. The past layer cannot play significant role in majority of normal radio pulsars evolution because the fast magnetic field evolution is not seen in these objects \citep{Igoshev2019}. In our simulations we assume $\tau_Q = 200$~Myr.      
Different magnetic field decay curves are shown in Figure~\ref{f:mf}.


NSs cool down differently depending on their initial mass. Here we use our fits to cooling curves calculated by \cite{Shternin2011_cooling_curves} for several masses.   In our calculations we neglect additional heating due to decaying magnetic field, as we are dealing with normal radio pulsars, so the decay time scale is relatively long and magnetic field energy is relatively low. 
For NSs with the mass $1.32$~M$_\odot$ we use the following fit:
\begin{equation}
T_\mathrm{crust} (t) = 7.65 \times 10^8 \;\mathrm{K} \left(\frac{t}{1 \, \mathrm{year}}\right)^{-0.182} \exp \left(-\frac{t}{857\,\mathrm{kyr}}\right).    
\end{equation}
Here $T_\mathrm{crust}$ is the temperature below the heat blanket.
For NS with mass $1.25$~M$_\odot$ we use:
\begin{equation}
T_\mathrm{crust} (t) = 1.78 \times 10^9 \;\mathrm{K} \left(\frac{t}{1 \, \mathrm{year}}\right)^{-0.26} \exp \left(-\frac{t}{1143\,\mathrm{kyr}}\right)    .
\end{equation}

For NSs with the lowest mass of $1.1$~M$_\odot$  the internal temperature drops significantly around the age $\approx 30$~kyr. This behaviour is not well described by a fit in a similar form.  To obtain a correct description of cooling in this case we interpolate the numerical curve with a cubic spline. 


In order to describe behaviour of braking index and spin-down age, we solve the differential eq. (\ref{e:decay}) together with eq. (\ref{e:slowing}) 
with the initial condition $P(0) = P_0$. We estimate the spin-down age and braking index at each time interval using eq. (\ref{e:brk}). 

\begin{figure*}
    \centering
    \begin{minipage}{0.49\linewidth}
    \includegraphics[width=1.0\linewidth]{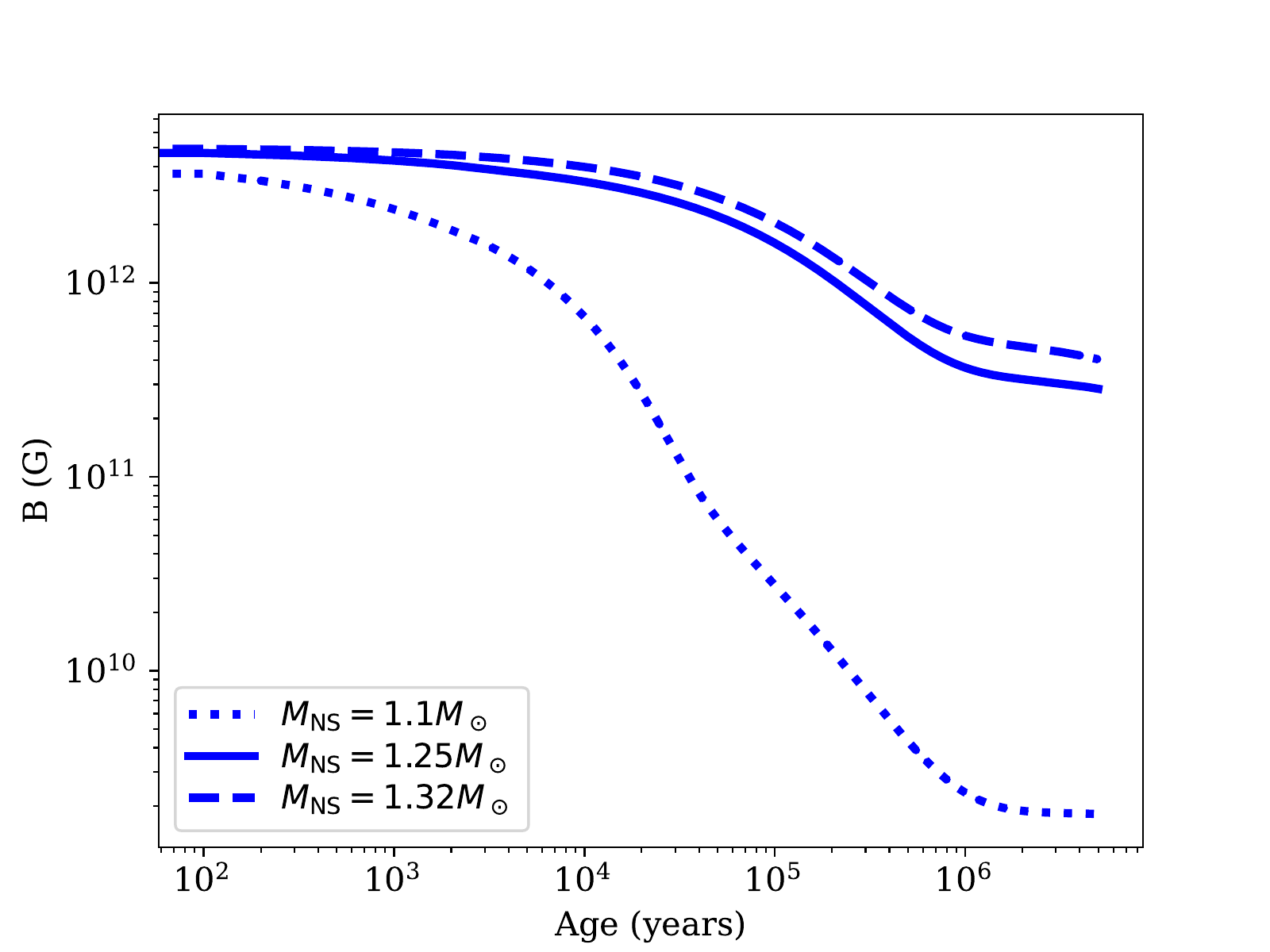}
    \end{minipage}
    \begin{minipage}{0.49\linewidth}
    \includegraphics[width=1.0\linewidth]{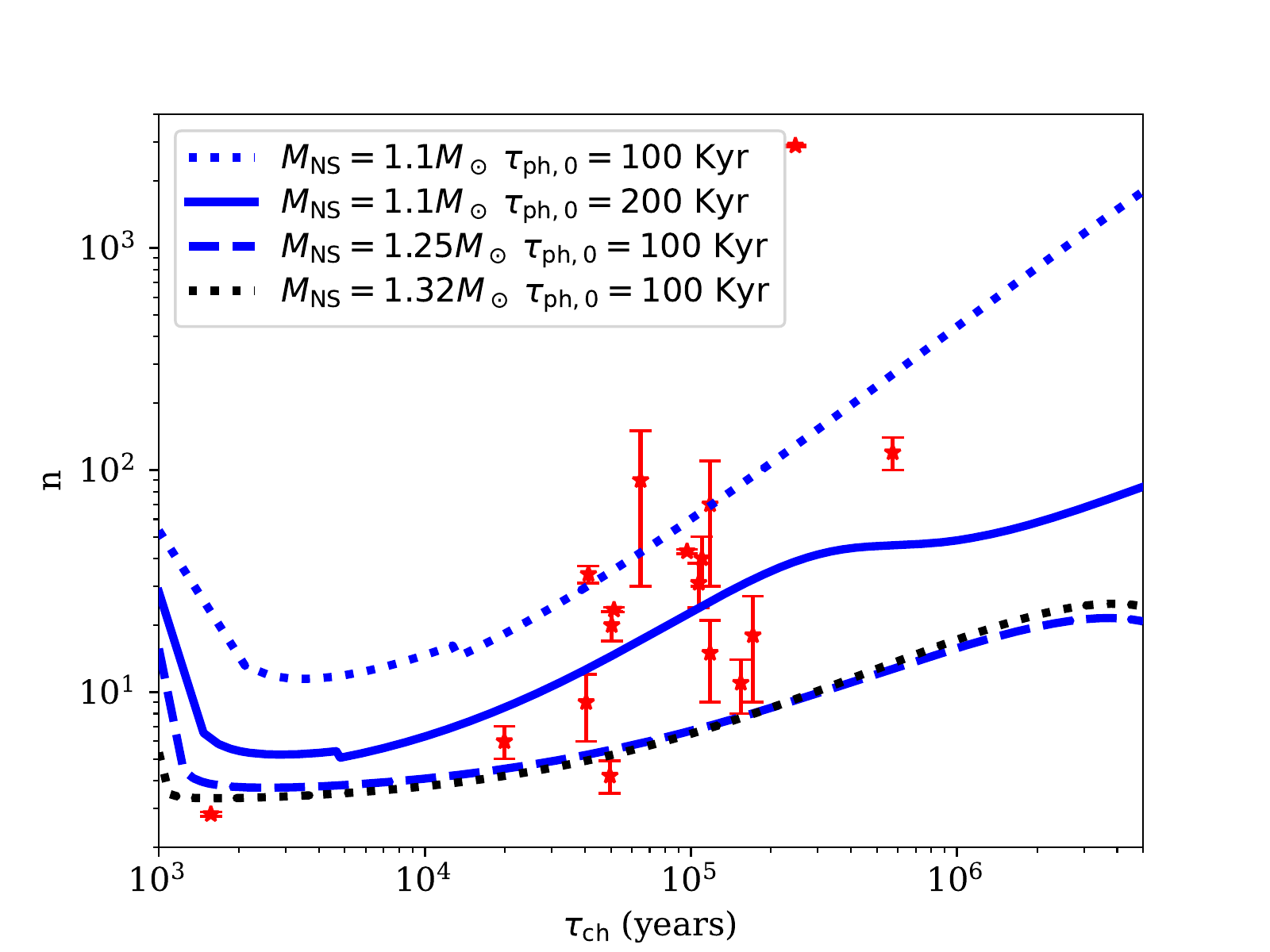}
    \end{minipage}    
    \caption{Left panel: magnetic field evolution for NSs of different masses, $\tau_\mathrm{ph,0}=10^5$~yrs and $\tau_Q = 200$~Myr. Right panel: dependence of braking index on spin-down age for different NS masses. Red dots show measured values for 17 isolated radio pulsars from \protect\cite{listBrakingInd}. We fix $P_0 = 0.04$~s.}
    \label{f:dm}
\end{figure*}

We show the dependence of braking indices on the spin-down ages in Figures~\ref{f:mf} and \ref{f:dm}. In the first plot we variate the initial period and typical phonon decay timescale, but keep the neutron star mass fixed at $M=1.32$~$M_\odot$. In Fig.~\ref{f:dm} we variate the NS mass (i.e., we modify the cooling curve, respectively) but keep the phonon decay timescale fixed at $\tau_\mathrm{ph,0} = 10^5$~yrs and initial period fixed at $P_0 = 0.04$~s. 
We see that $n(\tau_\mathrm{ch})$ curves reach values smaller than $n\approx 70$ before $\tau_\mathrm{ch}\approx$1~Myr if $M_\mathrm{NS} = 1.32$~M$_\odot$ even if unrealistically small $\tau_\mathrm{ph,0}$ are selected. On the other hand, in Figure~\ref{f:dm} we see that the curve for $M=1.1$~M$_\odot$ can explain large braking indices even if $\tau_\mathrm{ph,0} = 10^5$~yrs which is close to the realistic estimate. This result is in correspondence with findings in \cite{igoshev2014}, where we analysed pulsars with few$\times10^4\lesssim\tau_\mathrm{ch}<10^6$~yrs and demonstrated that $P-\dot P$ data require field decay with  $\tau\sim$few$\times10^5$~yrs which is switched-off for older objects. 

In the set of cooling curves that we use, the low-mass NSs cool down much slower than massive NSs. This is a general feature of many models of thermal evolution of NSs \citep{2004ARA&A..42..169Y, 2015SSRv..191..239P}. It happens because of much weaker neutrino emission as  bremsstrahlung in neutron-neutron collisions dominates in such stars down to their cores. Other much more efficient cooling mechanisms, even modified URCA process, are damped by proton superfluidity, and they become available only at higher densities, i.e., in more massive NSs \citep{2004ARA&A..42..169Y}. Therefore, internal temperature stays above $10^8$~K for a  much longer period of time which enhances the phonon resistivity. Phonon resistivity  results in significant decay of the magnetic field (up to a few orders of magnitude) which strongly increases the braking index. Thus, we can potentially ``weigh'' NSs by timing technique.
Of course, precise ``weighting'' requires much better knowledge of thermal and field evolution, as well as a better model of spin-down is necessary. Without such information we cannot determine precisely the masses necessary to explain a large observed value of the braking index.  Still, an optimistic conclusion is that we can obtain a model-dependent  guess of the fraction of low-mass NSs.

It is unclear whether there are enough low-mass young pulsars in the total radio pulsar population to explain observed braking indices.  Formation of such low-mass NSs was discussed in different scenarios, for example, in the ultra-stripped supernova model \citep{2015MNRAS.451.2123T}.  \cite{Antoniadis} analysed masses of millisecond radio pulsars which went through an episode of accretion.
If we directly use their bimodal model, we find that $1.1$~M$_\odot$ is 4.5$\sigma$ away from the peak, so a negligible number of radio pulsars could have this or lower mass. If we  use their unimodal model, we find that $1.1$~M$_\odot$ is 1.7$\sigma$ away from the peak, so up to $10$~per~cent of radio pulsars could be born with this or smaller mass. 
If the model of \cite{Antoniadis} is applicable only for millisecond pulsars,  the isolated radio pulsar with $M_\mathrm{NS}\approx 1.1$~M$_\odot$ might be more frequent. In particular, if a typical millisecond pulsar accretes $\approx 0.1$~M$_\odot$, the peak of the mass distribution by \cite{Antoniadis} will shift to smaller masses. 
Note that the population of millisecond pulsars might be different from isolated NSs due to selection effects as well. Thus, considerations about the mass distribution of isolated NSs may be just partly based on the data on recycled pulsars.

The main test for the enhanced phonon resistivity hypothesis is to measure NSs temperatures. This test is challenging for young radio pulsars since their X-ray emission is often dominated by bright non-thermal component. NSs with large braking indices are the ones which experience fast magnetic field decay, momentarily, so  even without accounting for additional heating, which, as we stated above, can be neglected in our model, their bulk surface thermal luminosity (and blackbody temperatures respectively) should be systematically larger than thermal luminosities and temperatures of pulsars with braking index $\approx 3$ and comparable spin-down ages,  as the later have systematically large masses. If fast magnetic field decay is indeed associated with the lowest mass neutron stars, these NSs typically have largest radii, additionally enhancing their surface thermal luminosity.

\subsection{Crystal impurities}

If  impurities in the crust are large in some normal radio pulsars and reach values of $Q = 20-200$ (similar to magnetars \citealt{Pons2013}), it also can cause braking indices $n\approx 10-100$. In comparison to the phonon decay, decay due to impurities is not terminated when the NS is getting cold enough. Therefore  radio pulsar activity of such NSs might cease on a  timescale not much longer than the one of the Ohmic decay, $\sim$a few $\times 10^5$~years. It seems impossible for a large group of radio pulsars to have impurities $Q\approx 100$, because the magnetic field would decay before pulsars reach $P\approx 1$~s and this decay should be easy to notice in analysis similar to \cite{Igoshev2019}.  

If one assumes that large values of $Q$ are responsible for large braking indices of some NSs, then these compact objects must rapidly loose their fields, and on time scale $\sim 10^6$~yrs we expect $B\sim 10^8$~G.  This possibility can also be tested with high-mass X-ray binaries (HMXBs). NSs in such systems spend $\sim$ few Myrs at the Ejector and Propeller stages prior to the moment when accretion from the secondary companion is started  (see \citealt{1992ans..book.....L} for the stages description). In addition, accretion rate from stellar wind is low, thus, in $\sim1$~Myr just $\sim10^{-4}\, M_\odot$ is accreted, which is smaller than the mass of the crust. Thus, magnetic field of such NSs have enough time to evolve down to low values in the regime equivalent or similar to the one in normal radio pulsars, if decay is driven by large impurities.  HMXBs with low-field NSs can be selected by short spin periods, as accreting compact objects are expected to reach approximate equilibrium between spin-up and spin-down processes. Analysis of known sources of this type demonstrates that there is no significant fraction of low-field NSs in HMXBs (see \citealt{2012NewA...17..594C, 2012ASPC..466..191P} and references therein). We conclude that  impurities in the crust are not a viable explanation for large values of braking index of noticeable fraction
of young pulsars.

\section{Evolution of obliquity angle}
\label{s:obliquity}

The angle between directions of the magnetic dipole and rotational axis of the radio pulsar is expected to evolve due to  electromagnetic forces caused by the magnetosphere (see, e.g, \citealt{philippov2014} for a recent discussion and references to previous studies).  Earlier analytical studies, which considered vacuum magnetospheres \citep{vacuum_case} predicted a fast alignment of the rotational axis with the dipole one. Such a configuration does not slow its rotation due to the electromagnetic emission and can reach arbitrary large values of the braking index. 

First analytical approaches in analysis of the plasma-filled magnetosphere \citep{1993ppm..book.....B} predicted that the evolution proceeded towards the counter-alignment.  
Modern detailed MHD studies of the filled magnetosphere show alignment of the magnetic dipole orientation and the rotational axis \citep{philippov2014}. This alignment proceeds much more gradually than in the case of vacuum magnetosphere. The configuration with small obliquity angle $\alpha$  still slows down effectively, which agrees with observations. Therefore, in plasma filled magnetosphere it is hard to reach large braking indices. Below we perform basic calculations to demonstrate quantitatively validity of these statements.

\subsection{Vacuum magnetosphere}

In a theoretical limit of empty magnetosphere \citep{vacuum_case} the pulsar slows down as:
\begin{equation}
P \dot P = \frac{2}{3} \beta B_p^2 \sin^2 \alpha    .
\end{equation}
The obliquity angle evolves with time as:
\begin{equation}
\sin \alpha (t) = \sin \alpha_0 \exp \left( - \frac{t}{\tau_v} \right),    
\end{equation}
where $\alpha_0$ is the initial obliquity angle and $\tau_v$ is the evolution timescale which is equal to:
\begin{equation}
\tau_v = \frac{3}{2} \frac{P_0^2}{\cos^2\alpha_0} \frac{1}{\beta B_p^2}   . 
\end{equation}
Note, that in this model {\it a priori} there are no specific restrictions on the value of $\tau_v$ because it is very sensitive to $\cos \alpha_0$, and from observations the distribution of initial obliquity angle is not well constrained. 
If we follow the procedure described in Section~\ref{s:simple_est} we obtain the following estimate:
\begin{equation}
n = 3 + 4 \frac{\tau_\mathrm{ch}}{\tau_v}    .
\end{equation}
To explain the large braking indices we require $\tau_v\approx 10^4-10^5$~yrs, which is not consistent with observational data. Observations indicate that the aligment proceeds on timescale of $\sim 7\times 10^7$~yrs \citep{Weltevrede2008}. Additionally, the vacuum magnetosphere struggles to explain formation of radio pulsars with periods of $\approx 1$~sec because aligned rotators in this model do not spin down significantly \citep{vacuum_case}.

\subsection{Plasma-filled magnetosphere}

\begin{figure*}
    \centering
    \begin{minipage}{0.49\linewidth}
    \includegraphics[width=1.0\linewidth]{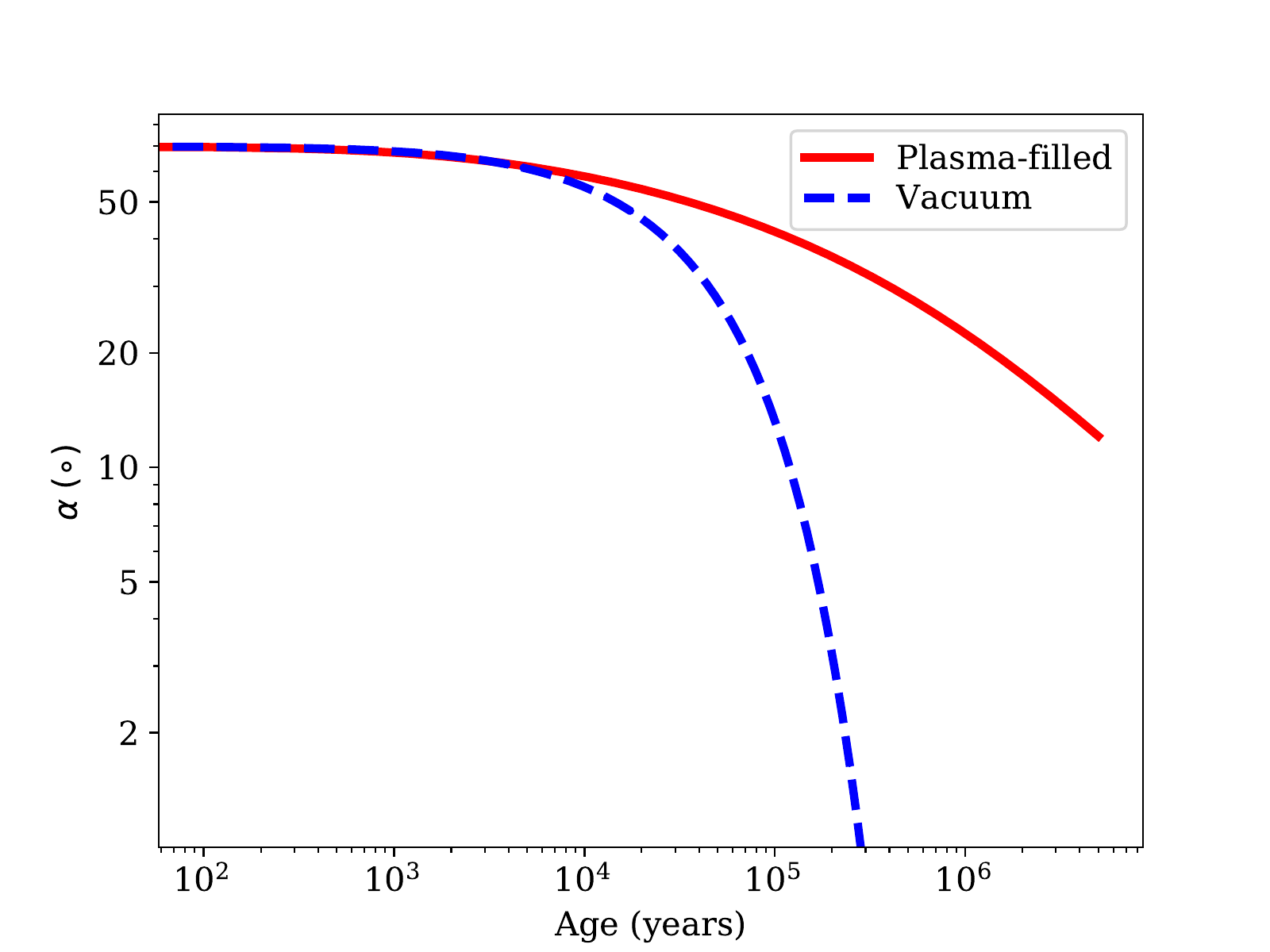}
    \end{minipage}
    \begin{minipage}{0.49\linewidth}
    \includegraphics[width=1.0\linewidth]{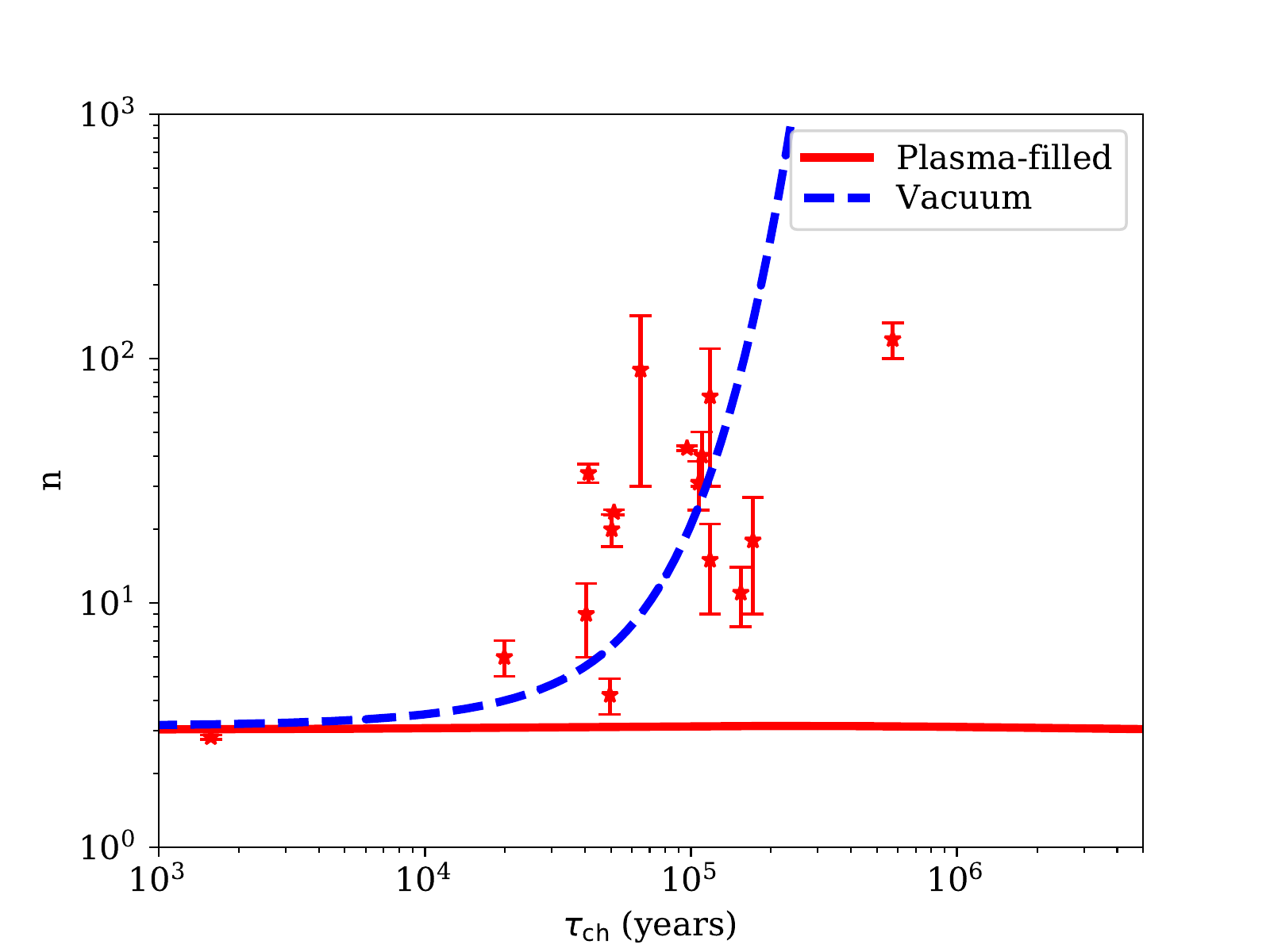}
    \end{minipage}    
    \caption{Left panel: Evolution of the obliquity angle as the function of time. Right panel: dependence of braking index on spin-down age for plasma-filled and vacuum magnetosheres of NSs. }
    \label{f:d_obl}
\end{figure*}

To compute the spin evolution of a pulsar with plasma-filled magnetosphere we solve simultaneously two equations:
\begin{equation}
P\dot P = (\kappa_0 + \kappa_1 \sin^2 \alpha) \beta B_p^2   
\end{equation}
and
\begin{equation}
\frac{d\alpha}{dt} = -\kappa_2 \beta \frac{B_p^2}{P^2} \sin \alpha \cos \alpha   .
\end{equation}
Here values of constants $\kappa_0 \approx 1$, $\kappa_1 \approx 1.2$ and $\kappa_2 \approx 1$  are estimated via MHD simulations by \cite{philippov2014} and $\alpha$ is the obliquity angle, as above. These equations are solved using the fourth-order Runge-Kutta integration method with initial conditions $P(0) = P_0$ and $\alpha (0) = \alpha_0$. 
As soon as the period, the period derivative, and the obliquity angle are computed we calculate the braking index using:
\begin{equation}
n = 3 - 2 \frac{\kappa_1 \sin 2 \alpha}{(\kappa_0 + \kappa_1 \sin^2 \alpha)} \tau_\mathrm{ch} \frac{d\alpha}{dt}    .
\end{equation}
This can be simplified further:
\begin{equation}
n = 3 + 2 \kappa_1\kappa_2 \frac{ \sin^2\alpha \cos^2\alpha}{(\kappa_0 + \kappa_1\sin^2\alpha)^2}    
\end{equation}
From this equation we can see \citep{Arzamasskiy2015} that the braking index in plasma filled magnetosphere varies between 3.0 and 3.25. The results of our calculations for both vacuum magnetosphere and plasma-filled magnetosphere are shown in Figure~\ref{f:d_obl} for $P_0 = 0.01$~s and $\alpha_0 = 70^\circ$ (left panel). We see that the obliquity angle reaches values which are very close to $0^\circ$ for vacuum magnetosphere. In this case the braking behaves as $\sin^2 \alpha$ which means that pulsars with obliquity angle close to 0 are not slowed down and the braking index can reach arbitrary large values, see Figure~\ref{f:d_obl} (right panel; for vacuum magnetosphere we use $P_0 = 0.04$~s). The presence of plasma in the pulsar magnetosphere changes this situation. In this case the angle evolves much slower. Also a pulsar with obliquity angle exactly $0^\circ$ is still slowed down and has braking index close to 3. The maximum braking index for the plasma filled magnetoshpere showed in Figure~\ref{f:d_obl} has a value of 3.125.

Overall, a pulsar with vacuum magnetosphere can reach arbitrary large braking indices at any moment of its evolution. The exact value is determined mostly by the initial obliquity angle. Vacuum magnetosphere is not considered to be physical because it contradicts observations. On the other hand the evolution of obliquity angle following the detailed MHD simulations shows that the braking index is always in range $n=3$~--~$3.25$ independent of the initial obliquity angle. Therefore, we do not think that obliquity angle evolution can explain braking indices measured by \cite{listBrakingInd}.


\section{Precession in the case of plasma filled magnetosphere}
\label{s:precession}

If the tensor describing the moment of inertia for radio pulsar $\hat I$ is not isotropic i.e. the mass distribution is not spherically symmetric and the rotation axis does not coincide with the axis of symmetry, the pulsar is expected to precess \citep{Melatos2000}. Precession means that the obliquity angle $\alpha$ evolves towards alignments as in the plasma filled magnetosphere model, but also oscillates \citep{Arzamasskiy2015}. 
These oscillations change $\dot P$ and $\ddot P$ altering the braking index significantly on timescales of $100-1000$~years.

Here we use following prescription \citep{Melatos2000, Arzamasskiy2015}: we project the angular rotational velocity $\Omega = 2\pi / P$ onto principal axes of the radio pulsar $\vec e_1, \vec e_2, \vec e_3$ using angles $\chi$ and $\theta$. Angle $\chi$ is the angle between dipole moment and axis $\vec e_3$. The angle $\theta$ is between the rotational axis and $\vec e_3$. In this representation, the inertia moment has components $I_1, I_2, I_3$. We simplify the problem and assume that $I_1 = I$, $I_2 = I (1 + \epsilon_{12})$ and  $I_3 = I (1 + \epsilon_{13})$. We simplify the problem further and consider a bi-axial star with $\epsilon_{12} = 0$.
 As soon as we choose initial conditions $\alpha_0$, $\chi_0$ and $\theta_0$ we calculate the component of the angular velocity as:
\begin{equation}
\omega_3 = \frac{2\pi}{P} \cos \theta_0 ,   
\end{equation}
\begin{equation}
\omega_1 = \frac{2\pi}{P} \frac{\cos \alpha_0 - \cos\theta_0 \cos \chi_0}{\sin \chi_0},    
\end{equation}
\begin{equation}
\omega_2 = \sqrt{\left(\frac{2\pi}{P}\right)^2 - \omega_3^2 - \omega_1^2}.    
\end{equation}

These values evolve further under influence of magnetospheric torques $\vec K$ following Euler's equations:
\begin{equation}
\hat I \frac{ d\vec \omega}{dt} + \vec \omega \times (\hat I \vec \omega) = \vec K    .
\end{equation}
This equation can be written in components as:
\begin{equation}
\frac{d\omega_1}{dt} = - \epsilon \omega_2 \omega_3 + \frac{K_1}{I},
\label{e:prec1}
\end{equation}
\begin{equation}
\frac{d\omega_2}{dt} = \epsilon \omega_3 \omega_1 + \frac{K_2}{I} ,   
\end{equation}
\begin{equation}
\frac{d\omega_3}{dt} =  \frac{K_3}{I (1+\epsilon)}   .
\label{e:prec3}
\end{equation}
The torques are taken from the MHD simulations by \cite{philippov2014}:
\begin{equation}
K_\mathrm{align} = \frac{B_p^2 \Omega^3 R^6_\mathrm{NS}}{0.25 c^3}    ,
\end{equation}
\begin{equation}
K_x = \kappa_2 K_\mathrm{align} \sin \alpha \cos \alpha   , 
\end{equation}
\begin{equation}
K_y = \kappa_3 K_\mathrm{align} \frac{c}{\Omega R_\mathrm{NS}}\sin \alpha \cos \alpha    ,
\end{equation}
\begin{equation}
K_z = - K_\mathrm{align} (\kappa_0 + \kappa_1 \sin^2 \alpha)    
\end{equation}
with numerical coeffitients $\kappa_0 = 1$, $\kappa_1 = 1.2$, $\kappa_2 = 1$ and $\kappa_3 = 0.1$. The vector of the torque is transformed using matrix $\vec K_{123} = \hat A \vec K_{xyz}$ to the coordinate system of the rotating body, see appendix 1 in \cite{Melatos2000}.
We solve the system of ordinary differential eqs. (\ref{e:prec1}-\ref{e:prec3}) using the fourth order Runge-Kutta integrator. At each numerical step we compute the obliquity angle as:
\begin{equation}
\alpha = \arccos \left(\frac{\omega_1 \sin \chi + \omega_3 \cos \chi}{\Omega}\right)  .  
\end{equation}
We perform a numerical simulation with $\theta_0 = 60^\circ$, $\chi = 4^\circ$, $\alpha=60^\circ$ and $\epsilon = 8\times 10^{-12}$.  We select parameters similar to \cite{Arzamasskiy2015} which they used to describe timing behaviour of the Crab radio pulsar with exception of $\chi$, which we increased slightly.

\begin{figure*}
    \centering
    \begin{minipage}{0.49\linewidth}
    \includegraphics[width=1.0\linewidth]{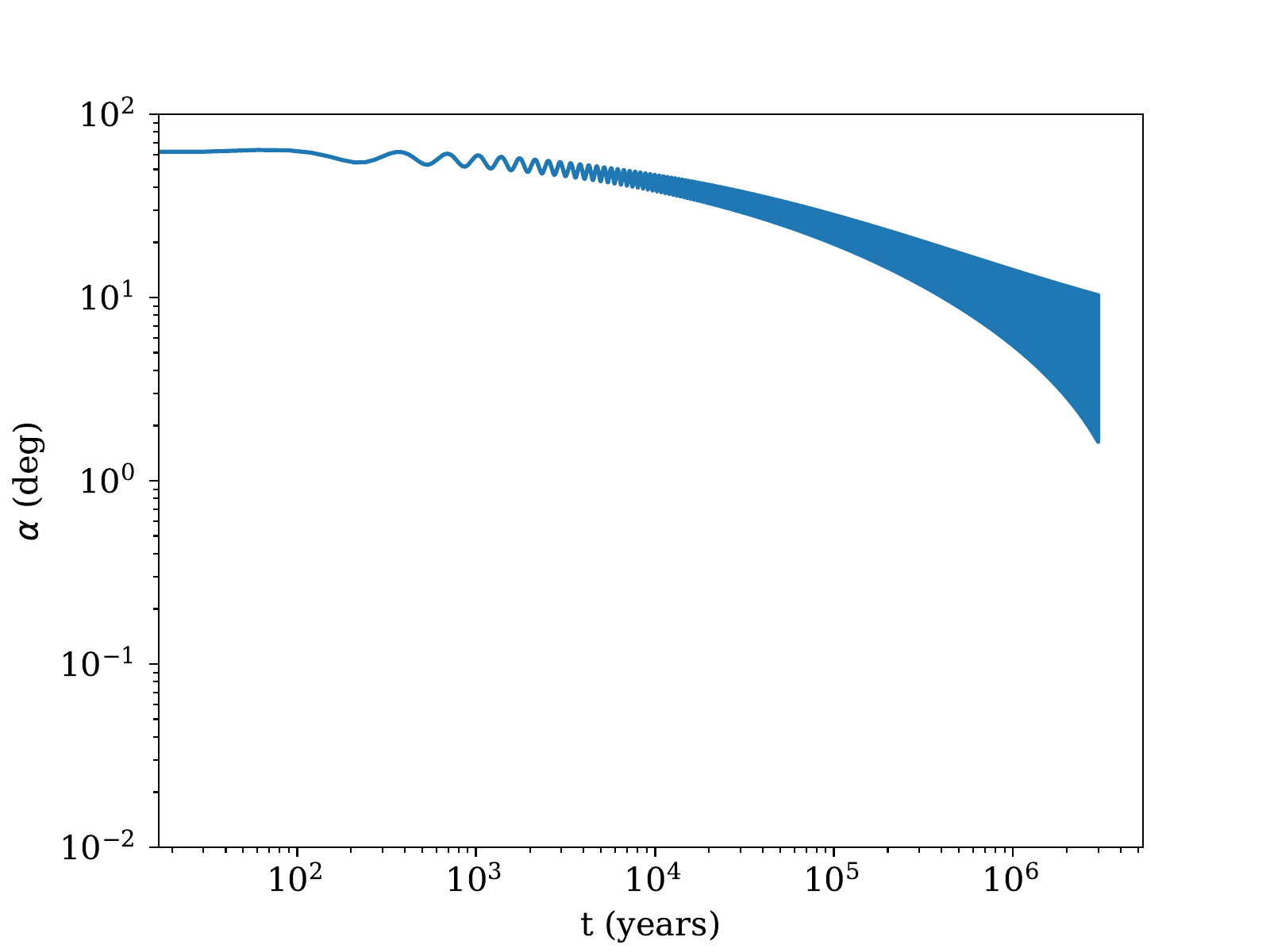}
    \end{minipage}
    \begin{minipage}{0.49\linewidth}
    \includegraphics[width=1.0\linewidth]{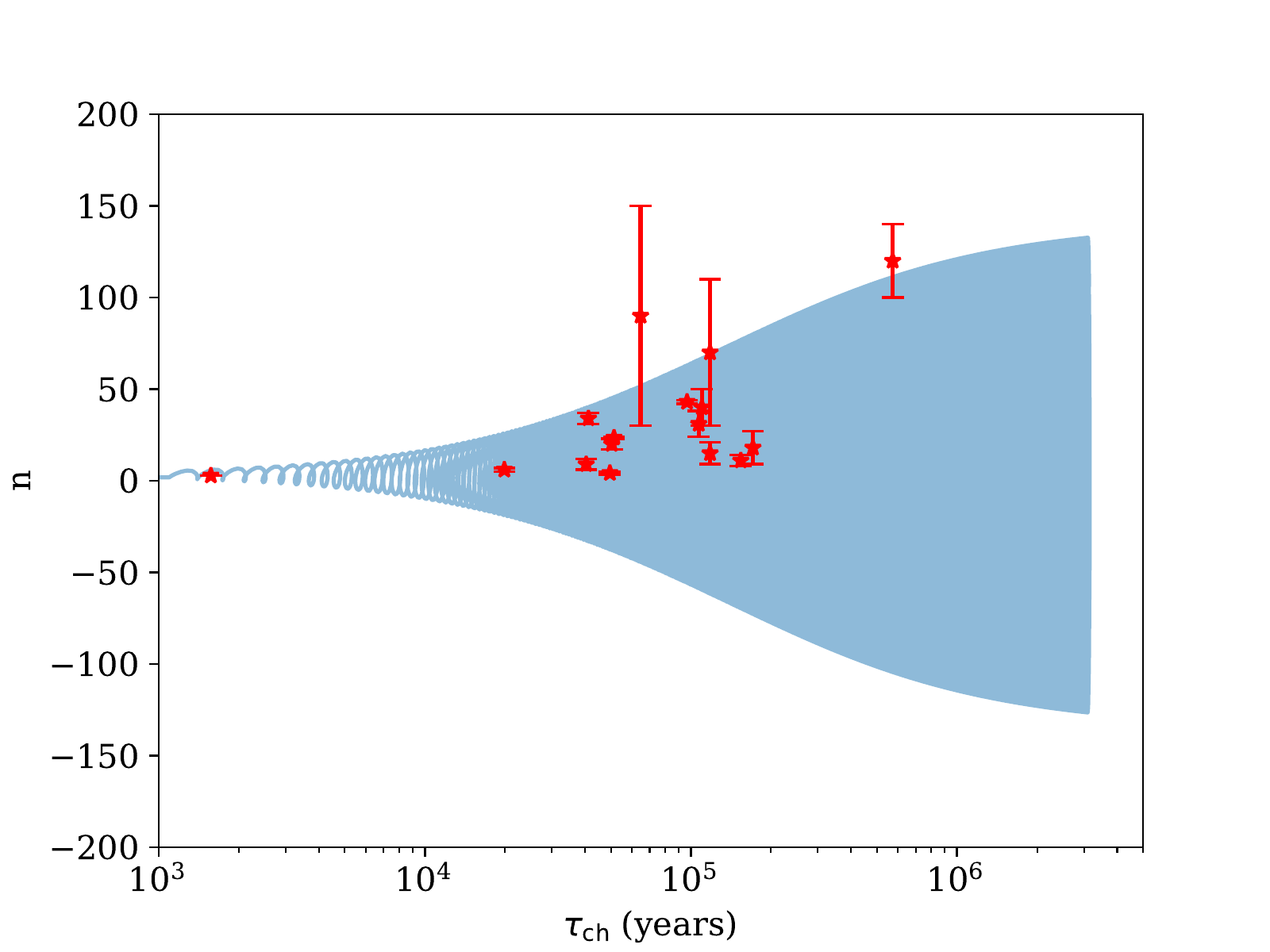}
    \end{minipage}    
    \caption{Left panel: evolution of the obliquity angle as the function of time. Right panel: dependence of braking index on spin-down age for plasma-filled magnetosphere with precession. }
    \label{f:d_precess}
\end{figure*}

We show results in Figure~\ref{f:d_precess}. It is easy to see that small oscillations of the obliquity angle causes the braking index to deviate significantly from $n=3$. These deviations are periodic with approximate period of:
\begin{equation}
T_\mathrm{prec} = \frac{P}{\epsilon}\approx 130~\mathrm{yrs}    .
\end{equation}
Half of the time the braking index is expected to be negative. It is seen in Figure~\ref{f:d_precess} (right panel) that a correlation is present between spin-down age and braking index as it requires some time for the precession to fully develop. Increase in the initial angle $\chi$ could produce even stronger modulation in the braking index.

Overall, the hypothesis that measured braking indices could be explained by rigid body precession has the same problems as the wide binary hypothesis: there are not enough negative braking indices. 
Hence, a fraction of the total sample might have non-standard braking indices because of precession.
If the extreme measured braking indices are indeed caused by pulsar precession this is easy to check in radio observations. The precession causes changes in the pulsar profile and the inclination geometry (probed though radio polarisation) should evolve with time.

\section{Non-dipolar magnetic field configurations}
\label{s:nondipole}

The braking index depends on configuration of the poloidal magnetic field of a NS. The braking index is most sensitive to the component of the field which stays dominant at the light-cylinder radius. \cite{krolik1991} showed that the braking index depends on the dominant magnetic multipole moment $l$ as $n=2l + 1$.
He also provided an equation to estimate the mean surface strength of non-dipolar component if it dominates:
\begin{equation}
\langle|B_{lm}|\rangle = S_{lm} \frac{(2l)!}{2^l l!} \sqrt{\frac{Ic^{2l+1}}{2^{2l-3}\pi^{2l-1}}}  \frac{1}{R_\mathrm{NS}^{l+2}} \sqrt{\dot P \left(\frac{P}{m}\right)^{2l-1}}, 
\end{equation}
where $S_{lm}$ is a numerical coefficient which is close to 1, exact values can be found in table 1 of \cite{krolik1991}.

At first, the dominant multipolar structure seems a plausible explanation 
for certain pulsars with not too large braking index. But if we develop this hypothesis further we notice the following.
 If we assume that pulsar J1815-1738 with braking index $n=9\pm 3$ has dominant octupole poloidal magnetic field $l=4$, it means that surface strength needs to be $\approx 10^{25}$~G to explain the combination of period and period derivatives, which is clearly nonphysical. The global magnetic field configuration 
 with pure hexapolar structure  ($l=3$) would require mean surface magnetic field $\approx 10^{20}$~G and could produce braking index of $n=7$. This strength is still nonphysical because the magnetic pressure would exceed gravity. 
 
It is worth to notice that pulsars with large braking indices do not look as outliers in the $P$~--~$\dot P$ diagram and do not demonstrate specific (for magnetars) types of activity. So it is hard to expect that these pulsars have much stronger fields.

\section{Conclusions}

In this article we analyse physical grounds for non-standard 
braking indices (\citealt{2019MNRAS.489.3810P}), $n\sim 10$ -- 100, of some young radio pulsars. At first, we restrict a number of possible wide binaries in the sample by \cite{listBrakingInd} based on the stellar statistics, a number of negative braking indices, and excess of optical counterparts seen in the Gaia second data release. We think that the majority of large braking indices (11/15) are intrinsic, thus they are not caused by  gravitational acceleration due to a stellar companion on a wide orbit. For these pulsars we investigated four hypotheses explaining  origin of large braking indices: (1) magnetic field decay, (2) evolution of the obliquity angle, (3) precession and (4) complicated multipole structure of the poloidal magnetic field.

We find that magnetic field decay is the only plausible explanation for the majority of large braking indices. The evolution of obliquity angle can cause large $n$ for certain initial obliquity angles only in the nonphysical case of vacuum magnetosphere. Plasma-filled magnetosphere gives $n$ ranges $3$~--~$3.25$ for all initial obliquity angles. Although  large multipoles $l=3,4$ can explain unusual braking indices of some objects, these surface fields need to have strength well in excess of physical limits for a NS stability. The precession can explain measured braking indices, but predicts equal number of positive and negative braking indices which is not observed.

Magnetic field decay can proceed with different speed in different NSs depending on crust composition (crust impurity parameter $Q$) or cooling of NSs. 
If in a large fraction of NSs the URCA processes are suppressed, the poloidal magnetic field could decay fast due to large phonon resistivity causing the braking index to reach values $n=10-100$. 

Overall, if the method suggested by \cite{2019MNRAS.489.3810P} does not discriminate against negative braking indices, the excess of large positive braking indices is statistically significant. Large positive braking indices can only be explained by magnetic field decay (possibly an episode of decay). Hence, it is necessary to study the phonon decay in details. Necessary prediction in this case is that large braking indices pulsars are hotter than normal braking indices objects. 
If the method of \cite{2019MNRAS.489.3810P} discriminates against negative braking indices, measurements of \cite{listBrakingInd} can be explained by a mixture of (1) presence of unseen companion (only a fraction of whole sample because of the stellar statistics) and (2) precession. Precession has simple observational signatures: evolution of pulsar profile and polarisation angle with time. Therefore, further observations of pulsars from the list of \cite{listBrakingInd} will be able to discriminate unambigiosly from these possibilities.

\section*{Acknowledgements}
We are grateful to the anonymous referee for comments and suggestions that helped to improve the manuscript.
SP thanks Peter Shternin for comments on cooling curves.
AI thanks support from the Science and Technology Facilities Council for research grant ST/S000275/1. SP acknowledges the support from the Program of development of M.V.Lomonosov Moscow State University (Leading Scientific School `Physics of stars, relativistic objects, and galaxies'). AI thanks A. Frantsuzova for help with proofreading of the manuscript.

This work has made use of data from the European Space Agency (ESA) mission
{\it Gaia} (\url{https://www.cosmos.esa.int/gaia}), processed by the {\it Gaia}
Data Processing and Analysis Consortium (DPAC,
\url{https://www.cosmos.esa.int/web/gaia/dpac/consortium}). Funding for the DPAC
has been provided by national institutions, in particular the institutions
participating in the {\it Gaia} Multilateral Agreement.

\section*{Data availability}
The data underlying this article will be shared on reasonable request to the corresponding author.





\bibliographystyle{mnras}
\bibliography{braking} 

\begin{thebibliography}{}
\makeatletter
\relax
\def\mn@urlcharsother{\let\do\@makeother \do\$\do\&\do\#\do\^\do\_\do\%\do\~}
\def\mn@doi{\begingroup\mn@urlcharsother \@ifnextchar [ {\mn@doi@}
  {\mn@doi@[]}}
\def\mn@doi@[#1]#2{\def\@tempa{#1}\ifx\@tempa\@empty \href
  {http://dx.doi.org/#2} {doi:#2}\else \href {http://dx.doi.org/#2} {#1}\fi
  \endgroup}
\def\mn@eprint#1#2{\mn@eprint@#1:#2::\@nil}
\def\mn@eprint@arXiv#1{\href {http://arxiv.org/abs/#1} {{\tt arXiv:#1}}}
\def\mn@eprint@dblp#1{\href {http://dblp.uni-trier.de/rec/bibtex/#1.xml}
  {dblp:#1}}
\def\mn@eprint@#1:#2:#3:#4\@nil{\def\@tempa {#1}\def\@tempb {#2}\def\@tempc
  {#3}\ifx \@tempc \@empty \let \@tempc \@tempb \let \@tempb \@tempa \fi \ifx
  \@tempb \@empty \def\@tempb {arXiv}\fi \@ifundefined
  {mn@eprint@\@tempb}{\@tempb:\@tempc}{\expandafter \expandafter \csname
  mn@eprint@\@tempb\endcsname \expandafter{\@tempc}}}

\bibitem[\protect\citeauthoryear{{Aguilera}, {Pons}  \& {Miralles}}{{Aguilera}
  et~al.}{2008}]{aguilera2008}
{Aguilera} D.~N.,  {Pons} J.~A.,   {Miralles} J.~A.,  2008, \mn@doi [\aap]
  {10.1051/0004-6361:20078786}, \href
  {https://ui.adsabs.harvard.edu/abs/2008A&A...486..255A} {486, 255}

\bibitem[\protect\citeauthoryear{{Antoniadis}, {Tauris}, {Ozel}, {Barr},
  {Champion}  \& {Freire}}{{Antoniadis} et~al.}{2016}]{Antoniadis}
{Antoniadis} J.,  {Tauris} T.~M.,  {Ozel} F.,  {Barr} E.,  {Champion} D.~J.,
  {Freire} P. C.~C.,  2016, arXiv e-prints, \href
  {https://ui.adsabs.harvard.edu/abs/2016arXiv160501665A} {p. arXiv:1605.01665}

\bibitem[\protect\citeauthoryear{{Arzamasskiy}, {Philippov}  \&
  {Tchekhovskoy}}{{Arzamasskiy} et~al.}{2015}]{Arzamasskiy2015}
{Arzamasskiy} L.,  {Philippov} A.,   {Tchekhovskoy} A.~e.,  2015, \mn@doi
  [\mnras] {10.1093/mnras/stv1818}, \href
  {https://ui.adsabs.harvard.edu/abs/2015MNRAS.453.3540A} {453, 3540}

\bibitem[\protect\citeauthoryear{{Beloborodov}}{{Beloborodov}}{2009}]{beloborodov2009}
{Beloborodov} A.~M.,  2009, \mn@doi [\apj] {10.1088/0004-637X/703/1/1044},
  \href {https://ui.adsabs.harvard.edu/abs/2009ApJ...703.1044B} {703, 1044}

\bibitem[\protect\citeauthoryear{{Beskin}}{{Beskin}}{2018}]{2018PhyU...61..353B}
{Beskin} V.~S.,  2018, \mn@doi [Physics Uspekhi] {10.3367/UFNe.2017.10.038216},
  \href {https://ui.adsabs.harvard.edu/abs/2018PhyU...61..353B} {61, 353}

\bibitem[\protect\citeauthoryear{{Beskin}, {Gurevich}  \& {Istomin}}{{Beskin}
  et~al.}{1993}]{1993ppm..book.....B}
{Beskin} V.~S.,  {Gurevich} A.~V.,   {Istomin} Y.~N.,  1993, {Physics of the
  pulsar magnetosphere}.
Cambridge University Press, Cambridge, UK

\bibitem[\protect\citeauthoryear{{Bilous} et~al.,}{{Bilous}
  et~al.}{2019}]{Bilous2019}
{Bilous} A.~V.,  et~al., 2019, \mn@doi [\apjl] {10.3847/2041-8213/ab53e7},
  \href {https://ui.adsabs.harvard.edu/abs/2019ApJ...887L..23B} {887, L23}

\bibitem[\protect\citeauthoryear{{Blandford} \& {Romani}}{{Blandford} \&
  {Romani}}{1988}]{1988MNRAS.234P..57B}
{Blandford} R.~D.,  {Romani} R.~W.,  1988, \mn@doi [\mnras]
  {10.1093/mnras/234.1.57P}, \href
  {https://ui.adsabs.harvard.edu/abs/1988MNRAS.234P..57B} {234, 57P}

\bibitem[\protect\citeauthoryear{{Braithwaite} \& {Spruit}}{{Braithwaite} \&
  {Spruit}}{2004}]{Braithwaite2004}
{Braithwaite} J.,  {Spruit} H.~C.,  2004, \mn@doi [\nat] {10.1038/nature02934},
  \href {https://ui.adsabs.harvard.edu/abs/2004Natur.431..819B} {431, 819}

\bibitem[\protect\citeauthoryear{{Chashkina} \& {Popov}}{{Chashkina} \&
  {Popov}}{2012}]{2012NewA...17..594C}
{Chashkina} A.,  {Popov} S.~B.,  2012, \mn@doi [\na]
  {10.1016/j.newast.2012.01.004}, \href
  {https://ui.adsabs.harvard.edu/abs/2012NewA...17..594C} {17, 594}

\bibitem[\protect\citeauthoryear{{Chen} \& {Beloborodov}}{{Chen} \&
  {Beloborodov}}{2017}]{2017ApJ...844..133C}
{Chen} A.~Y.,  {Beloborodov} A.~M.,  2017, \mn@doi [\apj]
  {10.3847/1538-4357/aa7a57}, \href
  {https://ui.adsabs.harvard.edu/abs/2017ApJ...844..133C} {844, 133}

\bibitem[\protect\citeauthoryear{{Chugunov}}{{Chugunov}}{2012}]{chugunov2012}
{Chugunov} A.~I.,  2012, \mn@doi [Astronomy Letters]
  {10.1134/S1063773712010021}, \href
  {https://ui.adsabs.harvard.edu/abs/2012AstL...38...25C} {38, 25}

\bibitem[\protect\citeauthoryear{{Cordes} \& {Lazio}}{{Cordes} \&
  {Lazio}}{2002}]{ne2001}
{Cordes} J.~M.,  {Lazio} T.~J.~W.,  2002, arXiv e-prints, \href
  {https://ui.adsabs.harvard.edu/abs/2002astro.ph..7156C} {pp
  astro--ph/0207156}

\bibitem[\protect\citeauthoryear{{Cumming}, {Arras}  \& {Zweibel}}{{Cumming}
  et~al.}{2004}]{cumming2004}
{Cumming} A.,  {Arras} P.,   {Zweibel} E.,  2004, \mn@doi [\apj]
  {10.1086/421324}, \href
  {https://ui.adsabs.harvard.edu/abs/2004ApJ...609..999C} {609, 999}

\bibitem[\protect\citeauthoryear{{Dall'Osso} \& {Perna}}{{Dall'Osso} \&
  {Perna}}{2017}]{2017MNRAS.472.2142D}
{Dall'Osso} S.,  {Perna} R.,  2017, \mn@doi [\mnras] {10.1093/mnras/stx2097},
  \href {https://ui.adsabs.harvard.edu/abs/2017MNRAS.472.2142D} {472, 2142}

\bibitem[\protect\citeauthoryear{{Deller} et~al.,}{{Deller}
  et~al.}{2019}]{deller2019}
{Deller} A.~T.,  et~al., 2019, \mn@doi [\apj] {10.3847/1538-4357/ab11c7}, \href
  {https://ui.adsabs.harvard.edu/abs/2019ApJ...875..100D} {875, 100}

\bibitem[\protect\citeauthoryear{{Faucher-Gigu{\`e}re} \&
  {Kaspi}}{{Faucher-Gigu{\`e}re} \& {Kaspi}}{2006}]{2006ApJ...643..332F}
{Faucher-Gigu{\`e}re} C.-A.,  {Kaspi} V.~M.,  2006, \mn@doi [\apj]
  {10.1086/501516}, \href
  {https://ui.adsabs.harvard.edu/abs/2006ApJ...643..332F} {643, 332}

\bibitem[\protect\citeauthoryear{{Ferrario}, {Melatos}  \& {Zrake}}{{Ferrario}
  et~al.}{2015}]{ferrario2015}
{Ferrario} L.,  {Melatos} A.,   {Zrake} J.,  2015, \mn@doi [\ssr]
  {10.1007/s11214-015-0138-y}, \href
  {https://ui.adsabs.harvard.edu/abs/2015SSRv..191...77F} {191, 77}

\bibitem[\protect\citeauthoryear{{Gaia Collaboration} et~al.,}{{Gaia
  Collaboration} et~al.}{2016}]{gaiaI}
{Gaia Collaboration} et~al., 2016, \mn@doi [\aap]
  {10.1051/0004-6361/201629272}, \href
  {https://ui.adsabs.harvard.edu/abs/2016A&A...595A...1G} {595, A1}

\bibitem[\protect\citeauthoryear{{Gaia Collaboration} et~al.,}{{Gaia
  Collaboration} et~al.}{2018}]{gaiaII}
{Gaia Collaboration} et~al., 2018, \mn@doi [\aap]
  {10.1051/0004-6361/201833051}, \href
  {https://ui.adsabs.harvard.edu/abs/2018A&A...616A...1G} {616, A1}

\bibitem[\protect\citeauthoryear{{Gonthier}, {Van Guilder}  \&
  {Harding}}{{Gonthier} et~al.}{2004}]{2004ApJ...604..775G}
{Gonthier} P.~L.,  {Van Guilder} R.,   {Harding} A.~K.,  2004, \mn@doi [\apj]
  {10.1086/382070}, \href
  {https://ui.adsabs.harvard.edu/abs/2004ApJ...604..775G} {604, 775}

\bibitem[\protect\citeauthoryear{{Gourgouliatos} \& {Cumming}}{{Gourgouliatos}
  \& {Cumming}}{2014}]{2014PhRvL.112q1101G}
{Gourgouliatos} K.~N.,  {Cumming} A.,  2014, \mn@doi [\prl]
  {10.1103/PhysRevLett.112.171101}, \href
  {https://ui.adsabs.harvard.edu/abs/2014PhRvL.112q1101G} {112, 171101}

\bibitem[\protect\citeauthoryear{{Gourgouliatos}, {Hollerbach}  \&
  {Archibald}}{{Gourgouliatos} et~al.}{2018}]{2018A&G....59e5.37G}
{Gourgouliatos} K.~N.,  {Hollerbach} R.,   {Archibald} R.~F.,  2018, \mn@doi
  [Astronomy and Geophysics] {10.1093/astrogeo/aty235}, \href
  {https://ui.adsabs.harvard.edu/abs/2018A&G....59e5.37G} {59, 5.37}

\bibitem[\protect\citeauthoryear{{Gourgouliatos}, {Hollerbach}  \&
  {Igoshev}}{{Gourgouliatos} et~al.}{2020}]{cco_3d}
{Gourgouliatos} K.~N.,  {Hollerbach} R.,   {Igoshev} A.~P.,  2020, \mn@doi
  [\mnras] {10.1093/mnras/staa1295}, \href
  {https://ui.adsabs.harvard.edu/abs/2020MNRAS.495.1692G} {495, 1692}

\bibitem[\protect\citeauthoryear{{Gull{\'o}n}, {Pons}, {Miralles},
  {Vigan{\`o}}, {Rea}  \& {Perna}}{{Gull{\'o}n}
  et~al.}{2015}]{2015MNRAS.454..615G}
{Gull{\'o}n} M.,  {Pons} J.~A.,  {Miralles} J.~A.,  {Vigan{\`o}} D.,  {Rea} N.,
    {Perna} R.,  2015, \mn@doi [\mnras] {10.1093/mnras/stv1644}, \href
  {https://ui.adsabs.harvard.edu/abs/2015MNRAS.454..615G} {454, 615}

\bibitem[\protect\citeauthoryear{{Hills}}{{Hills}}{1983}]{Hills1983}
{Hills} J.~G.,  1983, \mn@doi [\apj] {10.1086/160871}, \href
  {https://ui.adsabs.harvard.edu/abs/1983ApJ...267..322H} {267, 322}

\bibitem[\protect\citeauthoryear{{Igoshev}}{{Igoshev}}{2019}]{Igoshev2019}
{Igoshev} A.~P.,  2019, \mn@doi [\mnras] {10.1093/mnras/sty2945}, \href
  {https://ui.adsabs.harvard.edu/abs/2019MNRAS.482.3415I} {482, 3415}

\bibitem[\protect\citeauthoryear{{Igoshev}}{{Igoshev}}{2020}]{natalkickII}
{Igoshev} A.~P.,  2020, \mn@doi [\mnras] {10.1093/mnras/staa958}, \href
  {https://ui.adsabs.harvard.edu/abs/2020MNRAS.494.3663I} {494, 3663}

\bibitem[\protect\citeauthoryear{{Igoshev} \& {Popov}}{{Igoshev} \&
  {Popov}}{2014}]{igoshev2014}
{Igoshev} A.~P.,  {Popov} S.~B.,  2014, \mn@doi [\mnras]
  {10.1093/mnras/stu1496}, \href
  {https://ui.adsabs.harvard.edu/abs/2014MNRAS.444.1066I} {444, 1066}

\bibitem[\protect\citeauthoryear{{Igoshev} \& {Popov}}{{Igoshev} \&
  {Popov}}{2015}]{IgoshevPopov2015}
{Igoshev} A.~P.,  {Popov} S.~B.,  2015, \mn@doi [Astronomische Nachrichten]
  {10.1002/asna.201512232}, \href
  {https://ui.adsabs.harvard.edu/abs/2015AN....336..831I} {336, 831}

\bibitem[\protect\citeauthoryear{{Igoshev}, {Elfritz}  \& {Popov}}{{Igoshev}
  et~al.}{2016}]{2016MNRAS.462.3689I}
{Igoshev} A.~P.,  {Elfritz} J.~G.,   {Popov} S.~B.,  2016, \mn@doi [\mnras]
  {10.1093/mnras/stw1902}, \href
  {https://ui.adsabs.harvard.edu/abs/2016MNRAS.462.3689I} {462, 3689}

\bibitem[\protect\citeauthoryear{{Kochanek}, {Auchettl}  \&
  {Belczynski}}{{Kochanek} et~al.}{2019}]{2019MNRAS.485.5394K}
{Kochanek} C.~S.,  {Auchettl} K.,   {Belczynski} K.,  2019, \mn@doi [\mnras]
  {10.1093/mnras/stz717}, \href
  {https://ui.adsabs.harvard.edu/abs/2019MNRAS.485.5394K} {485, 5394}

\bibitem[\protect\citeauthoryear{{Krolik}}{{Krolik}}{1991}]{krolik1991}
{Krolik} J.~H.,  1991, \mn@doi [\apjl] {10.1086/186053}, \href
  {https://ui.adsabs.harvard.edu/abs/1991ApJ...373L..69K} {373, L69}

\bibitem[\protect\citeauthoryear{{Lipunov}}{{Lipunov}}{1992}]{1992ans..book.....L}
{Lipunov} V.~M.,  1992, {Astrophysics of Neutron Stars}.
Springer-Verlag, Berlin Heidelberg

\bibitem[\protect\citeauthoryear{{Melatos}}{{Melatos}}{2000}]{Melatos2000}
{Melatos} A.,  2000, \mn@doi [\mnras] {10.1046/j.1365-8711.2000.03031.x}, \href
  {https://ui.adsabs.harvard.edu/abs/2000MNRAS.313..217M} {313, 217}

\bibitem[\protect\citeauthoryear{{Mereghetti}}{{Mereghetti}}{2008}]{2008A&ARv..15..225M}
{Mereghetti} S.,  2008, \mn@doi [\aapr] {10.1007/s00159-008-0011-z}, \href
  {https://ui.adsabs.harvard.edu/abs/2008A&ARv..15..225M} {15, 225}

\bibitem[\protect\citeauthoryear{{Michel} \& {Goldwire}}{{Michel} \&
  {Goldwire}}{1970}]{vacuum_case}
{Michel} F.~C.,  {Goldwire} H.~C. J.,  1970, \aplett, \href
  {https://ui.adsabs.harvard.edu/abs/1970ApL.....5...21M} {5, 21}

\bibitem[\protect\citeauthoryear{{Moe} \& {Di Stefano}}{{Moe} \& {Di
  Stefano}}{2017}]{moe2017}
{Moe} M.,  {Di Stefano} R.,  2017, \mn@doi [\apjs] {10.3847/1538-4365/aa6fb6},
  \href {https://ui.adsabs.harvard.edu/abs/2017ApJS..230...15M} {230, 15}

\bibitem[\protect\citeauthoryear{{Ostriker} \& {Gunn}}{{Ostriker} \&
  {Gunn}}{1969}]{magnetodipole}
{Ostriker} J.~P.,  {Gunn} J.~E.,  1969, \mn@doi [\apj] {10.1086/150160}, \href
  {https://ui.adsabs.harvard.edu/abs/1969ApJ...157.1395O} {157, 1395}

\bibitem[\protect\citeauthoryear{{Page}, {Lattimer}, {Prakash}  \&
  {Steiner}}{{Page} et~al.}{2004}]{2004ApJS..155..623P}
{Page} D.,  {Lattimer} J.~M.,  {Prakash} M.,   {Steiner} A.~W.,  2004, \mn@doi
  [\apjs] {10.1086/424844}, \href
  {https://ui.adsabs.harvard.edu/abs/2004ApJS..155..623P} {155, 623}

\bibitem[\protect\citeauthoryear{{Parfrey}, {Beloborodov}  \& {Hui}}{{Parfrey}
  et~al.}{2012}]{2012ApJ...754L..12P}
{Parfrey} K.,  {Beloborodov} A.~M.,   {Hui} L.,  2012, \mn@doi [\apjl]
  {10.1088/2041-8205/754/1/L12}, \href
  {https://ui.adsabs.harvard.edu/abs/2012ApJ...754L..12P} {754, L12}

\bibitem[\protect\citeauthoryear{{Parthasarathy} et~al.,}{{Parthasarathy}
  et~al.}{2019}]{2019MNRAS.489.3810P}
{Parthasarathy} A.,  et~al., 2019, \mn@doi [\mnras] {10.1093/mnras/stz2383},
  \href {https://ui.adsabs.harvard.edu/abs/2019MNRAS.489.3810P} {489, 3810}

\bibitem[\protect\citeauthoryear{{Parthasarathy} et~al.,}{{Parthasarathy}
  et~al.}{2020}]{listBrakingInd}
{Parthasarathy} A.,  et~al., 2020, \mn@doi [\mnras] {10.1093/mnras/staa882},
  \href {https://ui.adsabs.harvard.edu/abs/2020MNRAS.494.2012P} {494, 2012}

\bibitem[\protect\citeauthoryear{{P{\'e}tri}}{{P{\'e}tri}}{2015}]{petri_multipoles}
{P{\'e}tri} J.,  2015, \mn@doi [\mnras] {10.1093/mnras/stv598}, \href
  {https://ui.adsabs.harvard.edu/abs/2015MNRAS.450..714P} {450, 714}

\bibitem[\protect\citeauthoryear{{Philippov}, {Tchekhovskoy}  \&
  {Li}}{{Philippov} et~al.}{2014}]{philippov2014}
{Philippov} A.,  {Tchekhovskoy} A.,   {Li} J.~G.,  2014, \mn@doi [\mnras]
  {10.1093/mnras/stu591}, \href
  {https://ui.adsabs.harvard.edu/abs/2014MNRAS.441.1879P} {441, 1879}

\bibitem[\protect\citeauthoryear{{Pons} \& {Vigan{\`o}}}{{Pons} \&
  {Vigan{\`o}}}{2019}]{2019LRCA....5....3P}
{Pons} J.~A.,  {Vigan{\`o}} D.,  2019, \mn@doi [Living Reviews in Computational
  Astrophysics] {10.1007/s41115-019-0006-7}, \href
  {https://ui.adsabs.harvard.edu/abs/2019LRCA....5....3P} {5, 3}

\bibitem[\protect\citeauthoryear{{Pons}, {Vigan{\`o}}  \& {Rea}}{{Pons}
  et~al.}{2013}]{Pons2013}
{Pons} J.~A.,  {Vigan{\`o}} D.,   {Rea} N.,  2013, \mn@doi [Nature Physics]
  {10.1038/nphys2640}, \href
  {https://ui.adsabs.harvard.edu/abs/2013NatPh...9..431P} {9, 431}

\bibitem[\protect\citeauthoryear{{Popov} \& {Turolla}}{{Popov} \&
  {Turolla}}{2012}]{2012ASPC..466..191P}
{Popov} S.~B.,  {Turolla} R.,  2012, in {Lewandowski} W.,  {Maron} O.,
  {Kijak} J.,  eds,  Astronomical Society of the Pacific Conference Series Vol.
  466, Electromagnetic Radiation from Pulsars and Magnetars. p.~191 (\mn@eprint
  {arXiv} {1206.2819})

\bibitem[\protect\citeauthoryear{{Popov}, {Pons}, {Miralles}, {Boldin}  \&
  {Posselt}}{{Popov} et~al.}{2010}]{2010MNRAS.401.2675P}
{Popov} S.~B.,  {Pons} J.~A.,  {Miralles} J.~A.,  {Boldin} P.~A.,   {Posselt}
  B.,  2010, \mn@doi [\mnras] {10.1111/j.1365-2966.2009.15850.x}, \href
  {https://ui.adsabs.harvard.edu/abs/2010MNRAS.401.2675P} {401, 2675}

\bibitem[\protect\citeauthoryear{{Potekhin}, {Pons}  \& {Page}}{{Potekhin}
  et~al.}{2015}]{2015SSRv..191..239P}
{Potekhin} A.~Y.,  {Pons} J.~A.,   {Page} D.,  2015, \mn@doi [\ssr]
  {10.1007/s11214-015-0180-9}, \href
  {https://ui.adsabs.harvard.edu/abs/2015SSRv..191..239P} {191, 239}

\bibitem[\protect\citeauthoryear{{Shternin}, {Yakovlev}, {Heinke}, {Ho}  \&
  {Patnaude}}{{Shternin} et~al.}{2011}]{Shternin2011_cooling_curves}
{Shternin} P.~S.,  {Yakovlev} D.~G.,  {Heinke} C.~O.,  {Ho} W. C.~G.,
  {Patnaude} D.~J.,  2011, \mn@doi [\mnras] {10.1111/j.1745-3933.2011.01015.x},
  \href {https://ui.adsabs.harvard.edu/abs/2011MNRAS.412L.108S} {412, L108}

\bibitem[\protect\citeauthoryear{{Tauris} \& {Manchester}}{{Tauris} \&
  {Manchester}}{1998}]{tauris1998}
{Tauris} T.~M.,  {Manchester} R.~N.,  1998, \mn@doi [\mnras]
  {10.1046/j.1365-8711.1998.01369.x}, \href
  {https://ui.adsabs.harvard.edu/abs/1998MNRAS.298..625T} {298, 625}

\bibitem[\protect\citeauthoryear{{Tauris}, {Langer}  \&
  {Podsiadlowski}}{{Tauris} et~al.}{2015}]{2015MNRAS.451.2123T}
{Tauris} T.~M.,  {Langer} N.,   {Podsiadlowski} P.,  2015, \mn@doi [\mnras]
  {10.1093/mnras/stv990}, \href
  {https://ui.adsabs.harvard.edu/abs/2015MNRAS.451.2123T} {451, 2123}

\bibitem[\protect\citeauthoryear{{Verbunt}, {Igoshev}  \& {Cator}}{{Verbunt}
  et~al.}{2017}]{natalkickI}
{Verbunt} F.,  {Igoshev} A.,   {Cator} E.,  2017, \mn@doi [\aap]
  {10.1051/0004-6361/201731518}, \href
  {https://ui.adsabs.harvard.edu/abs/2017A&A...608A..57V} {608, A57}

\bibitem[\protect\citeauthoryear{{Weltevrede} \& {Johnston}}{{Weltevrede} \&
  {Johnston}}{2008}]{Weltevrede2008}
{Weltevrede} P.,  {Johnston} S.,  2008, \mn@doi [\mnras]
  {10.1111/j.1365-2966.2008.13382.x}, \href
  {https://ui.adsabs.harvard.edu/abs/2008MNRAS.387.1755W} {387, 1755}

\bibitem[\protect\citeauthoryear{{Yakovlev} \& {Pethick}}{{Yakovlev} \&
  {Pethick}}{2004}]{2004ARA&A..42..169Y}
{Yakovlev} D.~G.,  {Pethick} C.~J.,  2004, \mn@doi [\araa]
  {10.1146/annurev.astro.42.053102.134013}, \href
  {https://ui.adsabs.harvard.edu/abs/2004ARA&A..42..169Y} {42, 169}

\bibitem[\protect\citeauthoryear{{Yao}, {Manchester}  \& {Wang}}{{Yao}
  et~al.}{2017}]{ymw2017}
{Yao} J.~M.,  {Manchester} R.~N.,   {Wang} N.,  2017, \mn@doi [\apj]
  {10.3847/1538-4357/835/1/29}, \href
  {https://ui.adsabs.harvard.edu/abs/2017ApJ...835...29Y} {835, 29}

\makeatother
\end{thebibliography}








\bsp	
\label{lastpage}
\end{document}